# Myopic Optimality: why reinforcement learning portfolio management strategies lose money

*Yuming MA

September 3, 2025

Department of Industrial Engineering and Economics, School of Engineering, Institute of Science Tokyo, Ōokayama 2-12-1, Meguro-ku, Tokyo 152-8552 Japan E-mail: ma.y.bb08@m.isct.ac.jp

**Abstract**

Myopic optimization (MO) outperforms reinforcement learning (RL) in portfolio management: RL yields lower or negative returns, higher variance, larger costs, heavier CVaR, lower profitability, and greater model risk. We model execution/liquidation frictions with mark-to-market accounting. Using Malliavin calculus (Clark-Ocone/BEL), we derive policy gradients and risk shadow price, unifying HJB and KKT. This gives dual gap and convergence results: geometric MO vs. RL floors. We quantify phantom profit in RL via Malliavin policy-gradient contamination analysis and define a control-affects-dynamics (CAD) premium of RL indicating plausibly positive.

## 1 Introduction

In this article we consider dynamic portfolio optimization for the buyside and the sellside mainly focusing on action-independent market (AIM) dynamics and standard execution/liquidation frictions with cases in control-affects-dynamics (CAD) market condition attached. In this regime, we show that myopic optimization (MO)—a sequence of one-period convex programs solved online—is dynamically sufficient relative to reinforcement learning (RL). The key device is a Malliavin "risk shadow price" that renders the Euler–Lagrange/KKT conditions equivalent to the HJB stationarity, enabling explicit dual gaps and effort-to-optimality bounds.

Notwithstanding the vogue for RL trading in quantitative finance research—it may look truer to the market and more originality in methodology, yet the apparatus remains the chief beneficiary—the admissible record is tepid. Under out-of-sample tests with implementable frictions (commissions, spreads, depth/latency), gains tend to vanish. Results premised on frictionless execution or unconstrained liquidity attenuate to statistical noise once costs are reinstated. This regularity is documented in field surveys [40, 67, 33]. Our analysis explains this pattern in AIM settings: MO converges geometrically, while RL faces a non-vanishing variance floor; at matched effort, MO achieves strictly higher mean profit and loss (PnL), lower variance, lighter CVaR, and larger positivity probability. For market making, the sell-side earns the bid–ask; yet in [75], an RL market-maker run across ten securities produces PnL estimates statistically indistinguishable from zero under a Student t-test (df = 9, 5 ). On hedging, most RL work rephrases risk control rather than mints PnL. Two mandates exist: (1) options pricing; (2) portfolio management.

In dynamic replication for option pricing, the design goal is near-zero expected PnL with small dispersion. In discrete time with trading frictions, convex-risk or risk-adjusted formulations inevitably embed a cost term, so the canonical outcome is a negative mean PnL; algorithms such as Deep Hedging [10, 12] primarily shrink variance and tails around that cost-induced negative mean rather than engineer a positive drift. Methodologically, it is best viewed as an RL-flavored instance of amortized optimization: a neural policy $\pi_\theta$ is trained offline on simulated paths to minimize a risk functional of terminal hedging error under frictions, thereby learning a reusable, $\mathcal{O}(1)$-inference mapping from state to trading action. In a representative derivation, [47] show that for a random-walk log-price with proportional costs the expected wealth increment equals the negative of expected trading costs, $\mathbb{E}[\text{PnL}] = -\mathbb{E}[\text{costs}]$, which formalizes the intuition that any "performance" in such settings comes from dispersion control, not from a positive mean.

By contrast, portfolio management seeks risk-controlled profitability, while RL usually delivers merely risk mitigation. [15] minimize transaction-cost mean plus a multiple of its standard deviation—no profitability claimed—while [16] report negative PnL. [57] manages an options portfolio and sharpens VaR/CVaR via Gamma-hedging with ATM options, the means of tabulated PnL remain predominantly negative. [21] employ reinforcement learning to hedge short barrier-option positions; however, as their Fig. 2 shows no material edge over a myopic optimizer or a Black–Scholes–Merton (BSM) delta hedging strategy. For the same task, [80] introduce robust RL, a distributionally robust optimization (DRO) method, within a Wasserstein ball; this stabilizes tails under misspecification but still takes prices as given and, under comparable frictions and rebalancing clocks, does not consistently dominate well-tuned myopic controls. Related work, e.g., [39], likewise centers on robust objectives in BSM-type settings rather than "beat-the-tape" profits. Where "profitability" appears, it is sometimes imported by construction: in [16], client options arrive via a Poisson flow and "earned" transaction costs of clients are netted against broker's trading costs; feasibility is declared when the former exceeds the latter—a fee transfer, not alpha.

Myopic optimization (MO) treats dynamic programs as a sequence of one-period convex programs solved at each time $t$, whereas the RL objective embeds each action's influence on future states and conducts long-horizon credit assignment via a Bellman recursion or cross-time gradient propagation. In frictionless settings, myopic optimality holds under isoelastic preferences [73, 60, 59]. With proportional trading costs $\lambda > 0$, optimal policies exhibit no-trade wedges and partial adjustment; small-cost band width $\delta \propto \lambda^{\frac{1}{3}}$, welfare-loss $\propto \lambda^{\frac{2}{3}}$ [25, 43, 44, 26, 82]. In option-hedging problems, boundary-hitting time $\tau_\delta$ mapping implies an optimal rebalancing frequency $f^* \propto \lambda^{-\frac{2}{3}} \iff \mathbb{E}[\tau_\delta] \propto \lambda^{\frac{2}{3}}$ [79, 78]. Conditional Value at Risk (CVaR) type tail-risk admits a convex reformulation posed by Rockafellar–Uryasev [71, 77, 70]; constraints and evaluation follow martingale/duality arguments and information-relaxation bounds, yielding shadow-price/KKT characterizations and computable optimality gaps [23, 24, 41]. Sequential MO is then implemented by projected/proximal updates with online-convex-optimization-style regret guarantees [83].

## 2 Theory

Portfolio management seeks optimal trading strategies that maximize a stated objective function under explicit constraints, e.g., risk appetite, inventory limits, rebalancing frequency and trading costs.

**Definition 2.1.** Let $\left(\Omega, \mathcal{F}, \mathbb{P}; \{\mathcal{F}_t\}_{t \geq 0}\right)$ be a complete filtered probability space satisfying the usual conditions. Let $B = \left(B^1, B^2, \ldots, B^d\right)^\top$ be a $d$-dimensional $\{\mathcal{F}_t\}$-Brownian motion with quadratic variation $d\left\langle B, B^\top \right\rangle_t = R dt$, $R \in \mathbb{S}_{++}^d$. The market state $Y_t \in \mathbb{R}^M$ solves the Stratonovich SDE

$$\circ dY_t = V_0\left(t, Y_t\right) dt + V\left(t, Y_t\right) \circ dB_t \tag{2.1}$$

with vector fields $V_0 : \mathbb{R}_+ \times \mathbb{R}^M \to \mathbb{R}^M$, $V = (V_1, V_2, \ldots, V_d) : \mathbb{R}_+ \times \mathbb{R}^M \to \mathbb{R}^{M \times d}$ and $V_0, V_1, \ldots, V_d \in C_{loc}^\infty\left(\mathbb{R}_+ \times \mathbb{R}^M; \mathbb{R}^M\right)$.

For $U \in \mathbb{R}^{M \times d}$, define the double-contraction operator with $\mathcal{D}$ Jacobian operator (note: $D$ for Malliavin derivatives)

$$\mathcal{J}_V[U] \coloneqq (\mathcal{D}V) : U = \sum_{k=1}^d J_k U_{:k} \qquad (\mathcal{J}_V[U])_i = \sum_{j=1}^M \sum_{k=1}^d (\mathcal{D}V)_{ijk} U_{jk} \in \mathbb{R}^M \tag{2.2}$$

where $\mathcal{D}V = \left((\mathcal{D}V)_{ijk}\right) \in \mathbb{R}^{M \times M \times d}$ is the Jacobian tensor

$$\mathcal{D}V \coloneqq [J_1 \mid \cdots \mid J_k \mid \cdots \mid J_d] \quad where \quad J_k \coloneqq \frac{\partial V_{:k}}{\partial x^\top} \in \mathbb{R}^{M \times M} \tag{2.3}$$

then we have the Itô-Stratonovich conversion

$$\text{Strat} \leftarrow \text{Itô}: \quad V_0(t, Y_t) = \widetilde{V}_0(t, Y_t) - \frac{1}{2}\mathcal{J}_V[V(t, Y_t) R] \tag{2.4}$$

where $\widetilde{V}_0$ is the drift vector field in Itô form, which is usually stated in SDE models.

Let $\varphi_{u \leftarrow s}(x)$ be the Kunita stochastic flow [48] generated by the Stratonovich SDE above, and let $J_{u \leftarrow s} \coloneqq \partial_x \varphi_{u \leftarrow s}(Y_s)$ denote the spatial Jacobian of $\varphi_{u \leftarrow s}(Y_s)$ with $J_{s \leftarrow s} = I_M$. Set $A(t, x) \coloneqq V(t, x) R V(t, x)^\top \in \mathbb{S}_+^M$. For any $t \geq 0$ and horizon length $T > 0$, define Malliavin covariance by writing

$$\Gamma_{t,T} \coloneqq \int_t^{t+T} J_{t+T \leftarrow s} A(s, Y_s) J_{t+T \leftarrow s}^\top ds \tag{2.5}$$

Assume Hörmander's bracket condition, this non-degeneracy holds [64, 51].

$$\Gamma_{t,T} \succ 0 \quad \text{a.s.}, \qquad \mathbb{E}\left[\left\|\Gamma_{t,T}^{-1}\right\|^p \mid \mathcal{F}_t\right] < \infty \text{ for some } p > 1 \quad \text{a.s.} \tag{2.6}$$

NOTE (change of measure): Results are stated under $\mathbb{P}$, however, when pricing purpose is required, pass to risk-neutral measure $\mathbb{Q}$ via Girsanov theorem under Novikov-Kazamaki conditions.

$$\frac{d\mathbb{Q}}{d\mathbb{P}} = \exp\left(-\frac{1}{2}\int_0^T \|\lambda_s\|^2 ds - \int_0^T \lambda_s^\top dB_s\right) \quad (2.7)$$

The filtration/predictability remains unchanged, thus our conclusions stay intact—only expectations switch to $\mathbb{Q}$. Specially, if anticipating or Wiener-space shifts are required, e.g., one may invoke the Cameron-Martin-Maruyama-Girsanov formula.

**Definition 2.2.** Let $Y_t \in \mathbb{R}^M$ be the market state, $P_t \in \mathbb{R}_+^N$ be the tradable price vector. Define mid or reference price $m_t := m(Y_t)$, half-spread $s_t := s(Y_t)$, limit-order-book (LOB) depth $D_t := D(Y_t)$. Let $\phi = (\phi_t)_{t\in[0,T]} \in \mathbb{R}^N$ be a predictable position process with trading speed rate $\dot{\phi}_t := \frac{d\phi_t}{dt}$ and execution order size $q_t$. Let $\tau_{\text{fill}}, \tau_{\text{liq}} > 0$ denote, respectively, the order fill horizon and the position liquidation horizon. Define two tradable price operators:

1. Execution operator $\Psi$:

   the executable price $P_t^{\text{exec}} = \Psi\left(Y_t; \dot{\phi}_t, q_t, \tau_{\text{fill}}, \text{mode}\right)$ is

$$P_t^{\text{exec}} = \begin{cases} m_t + s_t \text{sgn}(\dot{\phi}_t) + \Xi_t \dot{\phi}_t + \left(K_{\tau_{\text{fill}}} * \dot{\phi}\right)_t + g\left(\frac{|q_t|}{D_t}\right) & \text{mode} = \text{taker} \\ a_t \mathbf{1}_{\{dN_t^{\text{ask}} > 0\}} + b_t \mathbf{1}_{\{dN_t^{\text{bid}} > 0\}} & \text{mode} = \text{maker} \end{cases} \quad (2.8)$$

   where $\Xi_t \succeq 0$ is the temporary impact matrix,

$$\left(K_{\tau_{\text{fill}}} * \dot{\phi}\right)_t := \int_{(t-\tau_{\text{fill}})^+}^t K_{\tau_{\text{fill}}}(t-u) \dot{\phi}_u du \quad (2.9)$$

   is the transient impact with kernel $K_{\tau_{\text{fill}}} \in L^1(\mathbb{R}_+^N)$, and $g: \mathbb{R}_+^N \to \mathbb{R}^N$ is a convex overflow/slippage penalty. In market-making (MM) mode, let $\delta_t^\pm$ be quotation offsets, then

$$a_t = m_t + \delta_t^{\text{ask}}, \qquad b_t = m_t + \delta_t^{\text{bid}}, \qquad d\phi_t = dN_t^{\text{bid}} - dN_t^{\text{ask}} \quad (2.10)$$

   with passive order fill intensities $\lambda_t^\pm = \Lambda^\pm(\delta_t^\pm, Y_t)$. Passive position change $d\phi_t$ is exogenous in MM case with unilateral passive volume increment $dN_t^\pm$.

2. Liquidation operator $\Phi$:

   the mark-to-market (MtM) prudent valuation price is

$$P_t^{\text{liq}} = \Phi(Y_t; \phi_t, \tau_{\text{liq}}) = m_t - s_t \text{sgn}(\phi_t) - H\left(\frac{|\phi_t|}{\text{ADV}_t}, \tau_{\text{liq}}\right) \quad (2.11)$$

   where $H: \mathbb{R}^M \times \mathbb{R}_+ \to \mathbb{R}_+^N$ is a convex block-liquidation discount, which increases in size, position to average-daily-volume (ADV) ratio in particular, and decreases in horizon $\tau_{\text{liq}}$.

**Regulatory/accounting note.**

- For prudential capital (Basel PruVal / CRR Art. 105 [31, 29, 30]), the bid–ask convention is an acceptable close-out basis; mid is allowed only with evidence of close-out at mid.
- For financial reporting (IFRS 13 / ASC 820 [42, 32]), fair value is an exit price within the spread; bid–ask marking is permitted but not required (mid-market is also acceptable when it best represents exit price).

**Definition 2.3.** Basel PruVal / CRR Art. 105 Let $\mathcal{B}_t \in \mathbb{R}$ denote the cash balance (realized PnL aggregator), $\mathcal{V}_t := \langle \phi_t, P_t^{\text{liq}} \rangle \in \mathbb{R}$ the portfolio mark-to-market (MtM) value, and $\mathcal{X}_t := \mathcal{V}_t + \mathcal{B}_t$ the accounting wealth. Set $x^+ := \max\{x, 0\}$ and $x^- := \max\{-x, 0\}$. Fix predictable nonnegative processes $r_t^+$ (credit rate), $r_t^-$ (funding rate), and $\tau_t^{\text{tax}}$ (proportional transaction-tax rate). Let $m_t$ be the market mid price. Define the cash running gain rate $G_t$ in continuous trading (taker mode):

$$G_t\left(\phi_t, \dot\phi_t; Y_t\right) := -P_t^{\text{exec}} \dot\phi_t - \text{HC}_t(\phi_t) + r_t^+ \mathcal{B}_t^+ - r_t^- \mathcal{B}_t^- - \tau_t^{\text{tax}} m_t \left\| \dot\phi_t \right\|_1 \quad (2.12)$$

so that

$$\circ d\mathcal{B}_t = G_t\left(\phi_t, \dot\phi_t; Y_t\right) dt$$
$$= -P_t^{\text{exec}} \dot\phi_t dt - \text{HC}_t(\phi_t) dt + r_t^+ \mathcal{B}_t^+ dt - r_t^- \mathcal{B}_t^- dt - \tau_t^{\text{tax}} m_t \left\| \dot\phi_t \right\|_1 dt \quad (2.13)$$

the prudent valued MtM dynamics

$$\circ d\mathcal{V}_t = \left\langle \phi_t, \circ dP_t^{\text{liq}} \right\rangle + \left\langle P_t^{\text{liq}}, d\phi_t \right\rangle \quad (2.14)$$

Hence the wealth dynamics decompose as

$$\circ d\mathcal{X}_t = \circ d\mathcal{V}_t + \circ d\mathcal{B}_t$$
$$= \underbrace{\left\langle \phi_t, \circ dP_t^{\text{liq}} \right\rangle}_{\text{MtM PnL}}$$
$$+ \underbrace{\left\langle P_t^{\text{liq}} - P_t^{\text{exec}}, \circ d\phi_t \right\rangle - \text{HC}_t(\phi_t) dt + r_t^+ \mathcal{B}_t^+ dt - r_t^- \mathcal{B}_t^- dt - \tau_t^{\text{tax}} m_t \|\dot\phi_t\|_1 dt}_{\text{Realized PnL}} \quad (2.15)$$

Here $\text{HC}_t : \mathbb{R}^N \to [0, \infty)$ is convex and predictable; a canonical specification is

$$\text{HC}_t(\phi_t) = c_t^{\text{lend}} \phi_t^- + c_t^{\text{hold}} |\phi_t| + \zeta_t \left(\text{Margin}_t(\phi_t) - \text{FreeCash}_t\right)^+ \quad (2.16)$$

with predictable nonnegative $c_t^{\text{lend}}, c_t^{\text{hold}}, \zeta_t$.

For maker mode, with quotes $a_t$ and $b_t$ by the quoting strategy and point fills $dN_t^{\text{ask}}, dN_t^{\text{bid}}$,

$$d\phi_t = dN_t^{\text{bid}} - dN_t^{\text{ask}} \quad (2.17)$$

$$dB_t = a_t dN_t^{\text{ask}} + b_t dN_t^{\text{bid}} - \text{HC}_t(\phi_t)\,dt + r_t^+ \mathcal{B}_t^+ dt - r_t^- \mathcal{B}_t^- dt \qquad (2.18)$$

where one can optionally add the trading tax term $-\tau_t^{\text{tax}} m_t \|\dot{\phi}_t\|_1 dt$ to 2.18, and

$$d\mathcal{V}_t = \left\langle P_{t-}^{\text{liq}}, d\phi_t \right\rangle + \Delta\phi_t \Delta P_t^{\text{liq}} \qquad (2.19)$$

$$d\mathcal{X}_t = d\mathcal{V}_t + d\mathcal{B}_t \qquad (2.20)$$

If $P_t^{\text{liq}}$ is continuous, then $\Delta P_t^{\text{liq}} = 0$ and the last jump term vanishes.

**Convention: continuous vs. jump calculus** We use Stratonovich $\circ d$ for integrals with respect to the continuous Brownian part. All jump terms (e.g., passive market-making fills $dN_t^\pm$) are written with Itô $d$. Product rules follow semimartingale calculus: for any càdlàg semimartingales $X, Y$

$$d(X_t Y_t) = X_{t-} dY_t + Y_{t-} dX_t + d\langle X, Y \rangle_t, \quad \Delta(X_t Y_t) = X_{t-}\Delta Y_t + Y_{t-}\Delta X_t + \Delta X_t \Delta Y_t$$

Here $X_{t-}$ is the left limit, $\Delta X_t := X_t - X_{t-}$, and $\langle X, Y \rangle_t$ is the quadratic covariation (with continuous part $\langle X, Y \rangle_t^c$). For continuous parts, $X_t \circ dY_t = X_t dY_t + \frac{1}{2} d\langle X, Y \rangle_t^c$.

**Definition 2.4.** A feasible trading strategy is a predictable process $\phi = (\phi_t)_{t \in [0,T]} \in \mathbb{R}^N$ with finite variation and

$$\mathbb{E}\left[\int_0^T \left(|\phi_t| + |\dot{\phi}_t|\right) dt\right] < \infty. \qquad (2.21)$$

such that constraints below holds:

1. Inventory and trading speed rate bounds

$$\underline{\phi}_t \leq \phi_t \leq \bar{\phi}_t, \qquad \left\|\dot{\phi}_t\right\| \leq \bar{\nu}_t \quad \text{a.e. } t. \qquad (2.22)$$

2. Funding/borrow/margin capacity

$$\text{Margin}_t(\phi_t) \leq \text{Avail}_t(\phi_t^-), \qquad \text{ShortAvail}_t(\phi_t^-) \leq \text{BorrowCap}_t \qquad (2.23)$$

3. Risk and liquidity budgets under risk appetite policy (any convex, predictable constraints)

$$\rho_t(\Delta \mathcal{X}_{t,t+\Delta t}) \leq \bar{r}_t, \qquad \frac{|\phi_t|}{\text{ADV}_t} \leq \bar{\pi}_t \qquad (2.24)$$

for a designated conditional risk functional $\rho_t \in \mathbb{R}$

4. Market-making controls (if trading strategy is run by a market-maker)

$$\delta_t^\pm \in \left[\underline{\delta}_t, \overline{\delta}_t\right] \qquad (2.25)$$

It is determined by market maker duty and exchange/broker standards, Service-Level Agreement (SLA) for fills $(\tau_{\text{fill}}, q_{\min})$, quote lifetime $\tau_{\text{quote}}$, cancel/replace limits $L_{\text{CR}}$.

5. Terminal inventory (if applicable)

$$\left\|\phi_T - \phi_t^{\text{target}}\right\| \leq \varepsilon_T \qquad (2.26)$$

The collection of all such predictable controls is the feasible set $\mathfrak{F}$. Controls in $\mathfrak{F}$, their integrability conditions in Def. 2.2, 2.3 are well defined.

**Definition 2.5.** Let $\rho_t(Z)$ be a dynamic convex risk measure acting on losses $Z$ from wealth paths $\mathcal{X}_t$ and its Stratonovich stochastic increment $\circ d\mathcal{X}_t$, satisfying monotonicity, cash-invariant and conditional convexity, with instances:

- CVaR/ES: $\rho_t(Z) := \inf_{m \in \mathbb{R}} \left\{ m + \frac{1}{1-\alpha} \mathbb{E}\left[(Z-m)^+ \mid \mathcal{F}_t\right] \right\}$;

- Entropic risk with risk-aversion $\gamma$: $\rho_t^\gamma(Z) := \frac{1}{\gamma} \log \mathbb{E}\left[\exp(\gamma Z) \mid \mathcal{F}_t\right]$;

- Expectile risk: $\rho_t^\tau(Z) := \text{Exp}_\tau(Z \mid \mathcal{F}_t)$;

- Wasserstein-robust variant: $\rho_t(Z) := \sup_{Q \in \mathbb{B}_\varepsilon^{\mathcal{W}}} \mathbb{E}^Q[Z]$, and

$$\mathbb{B}_\varepsilon^{\mathcal{W}} = \{Q : \mathcal{W}_p(Q, \mathbb{P}[\cdot \mid \mathcal{F}_T]) \leq \varepsilon\}$$

a conditional $\mathcal{W}_p$ ball around the reference law.

**Definition 2.6.** General Portfolio Optimization Objective Functional

Let $\phi$ be a predictable position process with trading speed rate $\dot{\phi}$, and let $\lambda^\rho$ be the risk-aversion coefficient. The general portfolio optimization problem is

$$\max_{\phi \in \mathfrak{F}} J(\phi)$$

with either of the following equivalent objectives.

1. **Path-additive objective**

$$\max_{\phi \in \mathfrak{F}} J(\phi) := \mathbb{E}\left[\int_0^T \left(g_t\left(\phi_t, \dot{\phi}_t; Y_t\right) - \lambda^\rho \rho_t\left(\mathcal{X}_{[0,t]}\right)\right) dt\right] \qquad (2.27)$$

2. **Terminal objective with dynamic convex risk**

$$\max_{\phi \in \mathfrak{F}} J(\phi) := \mathbb{E}\left[U(\mathcal{X}_T)\right] - \lambda^\rho \rho_T(\mathcal{X}_T) \qquad (2.28)$$

**Definition 2.7.** Summary of Generalized Portfolio Optimization Programming Problem

$$\max_{\phi \in \mathfrak{F}} J(\phi) = \mathbb{E}\left[\int_0^T \left(G_t\left(\phi_t, \dot{\phi}_t; Y_t\right) - \lambda^\rho \rho_t\left(\mathcal{X}_{[0,t]}\right)\right) dt\right]$$

s.t. constraints in Def.2.4

$$\circ d\mathcal{B}_t = G_t dt, \ \circ d\mathcal{V}_t = \phi_t \circ dP_t^{\text{liq}} + P_t^{\text{liq}} \circ d\phi_t \ \mathcal{X}_t = \mathcal{V}_t + \mathcal{B}_t \quad (2.29)$$

**Proposition 2.1.** *Well-posedness*

*Assume: U concave and upper semi-continuous; for a.e. t, $G_t$ is concave in $\left(\phi_t, \dot{\phi}_t\right)$; $\rho_t$ is a proper lower semi-continuous convex functional; $\mathfrak{F}$ is closed and convex in the predictable topology with uniform integrability. Then J is upper semicontinuous and admits at least one maximizer $\phi^* \in \mathfrak{F}$.*

*Proof.* Let $\mathfrak{F}$ be convex, closed in the predictable topology, uniformly integrable; $U$ concave upper semicontinuous; for a.e. $t$, $\mathcal{B}_t$ is concave in $\left(\phi_t, \dot{\phi}_t\right)$; $\rho_t$ proper lower semicontinuous convex. Take a maximizing sequence $\{\phi^n\} \subset \mathfrak{F}$. By Komlós subsequence theorem, there exists forward convex combinations $\bar{\phi}^n$ converging a.s. predictably to some $\phi_t^* \in \mathfrak{F}$. Applying Fatou's Lemma with upper semicontinuous of concave terms and lower semicontinuous of convex penalty yield

$$\limsup_{n \to \infty} J\left(\bar{\phi}^n\right) \leq J\left(\phi^*\right) \quad (2.30)$$

Hence $\phi^*$ maximizes $J$. □

**Definition 2.8.** Following notation in Def. 2.1, we define optimizer SDE in Stratonovich form $\circ d\theta_\tau$. Let training time $\tau \in [0, T^{\text{opt}}]$, stochastic optimizer parameters $\theta_\tau \in \Theta \subset \mathbb{R}^p$, and optimizer driver—a Brownian motion $W_\tau \in \mathbb{R}^{d^{\text{opt}}}$, then the optimizer SDE is

$$\circ d\theta_\tau = V_0^{\text{opt}}\left(\tau, \theta_\tau\right) d\tau + V^{\text{opt}}\left(\tau, \theta_\tau\right) \circ dW_\tau^{\text{opt}} \quad (2.31)$$

with Itô drift $\widetilde{V}_0^{\text{opt}} = V^{\text{opt}} + \frac{1}{2}\mathcal{J}_{V^{\text{opt}}}\left[V^{\text{opt}}\left(\tau, \theta_\tau\right) R^{\text{opt}}\right]$, and optimizer covariance $A_{\text{opt}}(\theta) = V^{\text{opt}}(\theta) R^{\text{opt}} \left(V^{\text{opt}}(\theta)\right)^\top$. For optimizer vector field, a canonical choice is a preconditioned gradient flow, with preconditioner matrix $G \in \mathbb{S}_{++}^p$:

$$V_0^{\text{opt}}\left(\tau, \theta_\tau\right) = -G\left(\theta_\tau\right)^{-1} \nabla_\theta J\left(\theta_\tau\right) \quad (2.32)$$

## 2.1 Malliavin/ Clark-Ocone Gradient and Myopic First-Order Optimality

We state a general Gâteaux differentiation result, identify the risk shadow price via Malliavin/Clark–Ocone, and derive KKT conditions that reveal a myopic structure.

**Definition 2.9.** Let $\mathcal{Z}$ denote either the path argument $\mathcal{X}_{[0,t]}$ in Eq. 2.27 or $\mathcal{X}_T$ in Eq. 2.28. Define the predictable risk shadow price as

$$\lambda_t^{\text{risk}} := \mathbb{E}_t\left[D_t \partial \rho(\mathcal{Z})\right] \quad (2.33)$$

where $D_t$ is the Malliavin derivative and $\partial \rho$ any measurable selection from the subdifferential of the dynamic convex risk measures.

**Theorem 2.1.** *Clark-Ocone gradient and KKT system with general convex risk*

Assume $G_t$ is $C^1$ in $\left(\phi, \dot{\phi}\right)$, transient impact is represented by Volterra adjoint form, and constraints of feasible set $\mathfrak{F}$ holds. For any feasible variation $\delta\phi$, the Fréchet derivative of $J$ at $\phi$ is

$$DJ(\phi)[\delta\phi] = \left.\frac{d}{d\varepsilon}J(\phi + \varepsilon\delta)\right|_{\varepsilon=0} = \mathbb{E}\left[\int_0^T \langle \mathcal{G}_t(\phi), \delta\phi_t\rangle dt\right] \quad (2.34)$$

with gradient kernel

$$\mathcal{G}_t(\phi) = \underbrace{\partial_\phi G_t}_{\text{inventory margin}} - \frac{d}{dt}\left(\underbrace{\partial_{\dot{\phi}} G_t}_{\text{execution margin}} + \underbrace{p_t}_{\text{transient impact adjoint}}\right) - \underbrace{\lambda_t^{\text{risk}}}_{\text{risk shadow price}} \quad (2.35)$$

where, by setting $I_t := \left(K_{\tau_{\text{fill}}} * \dot{\phi}\right)_t$,

$$\delta I_t = \int_{(t-\tau_{\text{fill}})^+}^t K_{\tau_{\text{fill}}}(t-u)\,\delta\dot{\phi}_u du, \qquad \int_0^T \partial_I G_t \delta I_t dt = \int_0^T p_t \delta\dot{\phi}_t dt$$

the Volterra adjoint from convolution-typed transient impact [36, 81] is

$$p_t := \int_t^{(t+\tau_{\text{fill}})\wedge T} K_{\tau_{\text{fill}}}(s-t)^\top \partial_I G_s ds, \qquad p_T = 0 \quad (2.36)$$

and the KKT conditions are first-order necessary for an optimal portfolio manangement strategy $\phi^* \in \mathfrak{F}$

$$\mathcal{G}_t(\phi^*) + \alpha_t - \beta_t + \sum_k \eta_{k,t} \nabla_\phi c_{k,t}(\phi^*) = 0, \qquad \alpha_t\left(\phi_t^* - \overline{\phi}_t\right) = \beta_t\left(\underline{\phi}_t - \phi_t^*\right) = 0 \quad (2.37)$$

where $c_{k,t}$ collect the active constraints in Def. 2.4 and $\alpha_t, \beta_t, \{\eta_{k,t}\} \geq 0$ are predictable multipliers.

*Proof.* Let $I_t := \left(K_{\tau_{\text{fill}}} * \dot{\phi}\right)_t$. For feasible variation $\phi^\varepsilon := \phi + \varepsilon\delta\phi$ with $\delta\phi \in \mathfrak{F}$, and $\delta\phi_0 = \delta\phi_T = 0$,

$$\delta I_t = \int_{(t-\tau_{\text{fill}})^+}^t K_{\tau_{\text{fill}}}(t-u)\,\delta\dot{\phi}_u du$$

By the standard Volterra adjoint identity (see Lemma 2.3 in [81] or [36]) gives its Volterra adjoint,

$$p_t := \int_t^{(t+\tau_{\text{fill}})\wedge T} K_{\tau_{\text{fill}}}(s-t)^\top \partial_I G_s ds, \qquad p_T = 0 \quad (2.38)$$

$$\int_0^T \partial_I G_t \delta I_t dt = \int_0^T p_t \delta \dot{\phi}_t dt \qquad (2.39)$$

From Def. 2.3 and the above,

$$\delta \mathcal{B}_T = \int_0^T \left( \partial_\phi G_t \delta \phi_t + \partial_{\dot{\phi}} G_t \delta \dot{\phi}_t + \partial_I G_t \delta I_t \right) dt \qquad (2.40)$$

Substituting 2.39 into 2.40 yields

$$\delta \mathcal{B}_T = \int_0^T \left( \partial_\phi G_t \delta \phi_t + \partial_{\dot{\phi}} G_t \delta \dot{\phi}_t + p_t \delta \dot{\phi}_t \right) dt \qquad (2.41)$$

Using $\delta \phi_0 = \delta \phi_T = 0$,

$$\int_0^T \left( \partial_{\dot{\phi}} G_t + p_t \right) \delta \dot{\phi}_t dt = - \int_0^T \frac{d}{dt} \left( \partial_{\dot{\phi}} G_t + p_t \right) \delta \phi_t dt \qquad (2.42)$$

then

$$d\mathcal{B}_T = \int_0^T \left( \partial_\phi G_t - \frac{d}{dt} \left( \partial_{\dot{\phi}} G_t + p_t \right) \right) \delta \phi_t dt \qquad (2.43)$$

Let $z$ be the risk argument in the objective, and $\xi \in \partial \rho(z)$ and $\xi \in L^2 \cap \mathbb{D}^{1,2}$ be any measurable subgradient. According to Clark-Ocone formula [64] or its extention for Lévy/Poisson processes—exogenous passive position change $\circ d\phi_t$ may follow a Lévy/Poisson process—see Thm.12.22 in [65]).

$$\xi = \mathbb{E}[\xi] + \int_0^T \mathbb{E}_t [D_t \xi \mid \mathcal{F}_t] dB_t \qquad (2.44)$$

Thus the Gâteaux variation of the risk penalty equals

$$\delta \left( \mathbb{E} \left[ \int_0^T \rho_t dt \right] \right) = \mathbb{E} \left[ \int_0^T \lambda_t^{\text{risk}} \delta \phi_t dt \right], \qquad \lambda_t^{\text{risk}} := \mathbb{E}_t [D_t \xi \mid \mathcal{F}_t] \qquad (2.45)$$

When $\xi = f(Y_T)$—or a smooth functional of $Y$—the Malliavin chain rule together with

$$D_s Y_t = J_{t \leftarrow s}(Y_s) V(s, Y_s) \mathbf{1}_{\{s \leq t\}} \qquad (2.46)$$

gives the vector filed representation of the risk shadow price in terminal case

$$\lambda_t^{\text{risk}} = \mathbb{E}_t \left[ V(t, Y_t)^\top J_{T \leftarrow t}^\top \nabla f(Y_T) \mid \mathcal{F}_t \right] \qquad (2.47)$$

For path dependent functionals, the analogous Fréchet–Malliavin representation holds with the kernel acting on $D_s Y_t$.

Moreover, under Hörmander Malliavin non-degeneracy in Def. 2.1, one also has the Bismut-Elworthy-Li (BEL) weight [34]. [8, 28, 76]:

$$\lambda_t^{\text{risk}} = \mathbb{E}_t [D_t \xi \mid \mathcal{F}_t] = \mathbb{E}_t [\xi H_{t,T} \mid \mathcal{F}_t] \qquad (2.48)$$

$$H_{t,T} := \delta \left( \mathbf{1}_{[t,t+T]} V(s, Y_s)^\top J_{t+T \leftarrow s}(Y)^\top \Gamma_{t,T}^{-1} J_{t+T \leftarrow s}(Y) V(s, Y_s) \right) \in \mathbb{R}^d \tag{2.49}$$

Specially, in strongly elliptic (inversible diffusion) cases, with $W_t := R^{-\frac{1}{2}} B_t$ and $\Sigma(t, x) := V(t, x) R^{\frac{1}{2}}$

$$H_{t,T} = \frac{1}{T} \int_0^T \left( (\Sigma(s, Y_s))^{-1} J_{s \leftarrow t}(Y) \Sigma(s, Y_s) \right)^\top dW_s \tag{2.50}$$

Combining 2.43 with $\lambda_t^{\text{risk}}$ either in Clark-Ocone or BEL form yields the Euler-Lagrange gradient kernel

$$\mathcal{G}_t(\phi) = \partial_\phi G_t - \frac{d}{dt}\left( \partial_{\dot\phi} G_t + p_t \right) - \lambda_t^{\text{risk}}, \qquad p_t = \int_t^{(t+\tau_{\text{fill}}) \wedge T} K_{\tau_{\text{fill}}}(s-t)^\top \partial_I G_s ds \tag{2.51}$$

and Fréchet derivative of $J$ at $\phi$ is

$$DJ(\phi)[\delta\phi] = \frac{d}{d\varepsilon} J(\phi + \varepsilon \delta) \bigg|_{\varepsilon=0} = \mathbb{E}\left[ \int_0^T \langle \mathcal{G}_t(\phi), \delta\phi_t \rangle dt \right] \tag{2.52}$$

Afterwards, by standard convex optimality—see Thm.11. in [72] and Ch.3 in [6] for predictable multipliers $\alpha_t, \beta_t, \{\eta_{k,t}\} \geq 0$—hence, first-order necessary conditions at an optimal portfolio management strategy $\phi^* \in \mathfrak{F}$ are

$$\mathcal{G}_t(\phi^*) + \alpha_t - \beta_t + \sum_k \eta_{k,t} \nabla_\phi c_{k,t}(\phi^*) = 0, \qquad \alpha_t\left(\phi_t^* - \overline{\phi}_t\right) = \beta_t \left(\underline{\phi}_t - \phi_t^*\right) = 0 \tag{2.53}$$

$\square$

**Corollary 2.1.** *Myopic no-trade wedge and linear/soft-threshold rebalancing*

Suppose $G_t$ separates as

$$G_t = \widetilde{\mu}_t \phi_t - \frac{1}{2} \phi_t^\top Q_t \phi_t - \lambda_t \left\| \dot\phi_t \right\|_1 - \frac{1}{2} \dot\phi_t^\top \Xi_t \dot\phi_t - \text{HC}_t(\phi_t) \tag{2.54}$$

with $Q_t \succeq 0, \Xi_t \succ 0$. Then the KKT system reduces pointwise in $t$ to optimal trade speed rate $\dot\phi_t^*$

$$\begin{cases} \left\| \widetilde{\mu}_t - Q_t \phi_t^* - \nabla_\phi \text{HC}_t(\phi_t^*) - \lambda_t^{\text{risk}} \right\|_\infty \leq \lambda_t & \Rightarrow \dot\phi_t^* = 0 \\ \dot\phi_t^* = \Xi_t^{-1}\left( \widetilde{\mu}_t - Q_t \phi_t^* - \nabla_\phi \text{HC}_t(\phi_t^*) - \lambda_t^{\text{risk}} - \lambda_t \text{sgn}\left(\dot\phi_t^*\right) \right), & \text{otherwise} \end{cases} \tag{2.55}$$

Hence the optimal policy $\left(\phi_t^*, \dot\phi_t^*\right)$ is myopic: a time-local linear/soft-threshold feedback in $(\phi_t, Y_t)$ and transient impact states. The same structure hold for market-making after replacing $\widetilde{\mu}_t$ by the spread revenue term and optimizing over $\delta_t^\pm$.

**Definition 2.10.** Reinforcement Learning formulation

RL and MO solves the same optimization problem in Def. 2.7 but with a different formulation of Markov decision process (MDP) in continuous time. Take the RL state $S_t := (Y_t, \phi_t, I_t)$, action $A_t := \dot{\phi}_t$ in the feasible set $\mathfrak{F}$. Let $r_t := G_t\left(\phi_t, \dot{\phi}_t; Y_t\right) - \rho_t\left(\mathcal{X}_{[0,t]}\right)$.

A RL policy $\pi : S_t \mapsto A_t$ is a predictable mapping, and $\pi$ can be either linear or non-linear (e.g., Q-learning, Neural Networks, Wasserstein Robust DRO, etc.). The RL objective is

$$\max_{\pi \in \mathfrak{F}} J(\pi) := \mathbb{E}\left[\int_0^T r_t^\pi dt\right] = J(\phi^\pi) \tag{2.56}$$

where $\phi^\pi$ is the portfolio position path induced by $\pi$.

**Lemma 2.1.** *Hamilton-Jacobi-Bellman (HJB)/KKT identity*

Let $\mathbb{V}^\pi(t, S_t)$ be the value process under RL policy $\pi$ for state $S_t = (Y_t, \phi_t, I_t)$, with the Hamiltonian

$$\mathbb{H}_t(a) := G_t(\phi_t, a; Y_t) - \rho_t + \langle p_t, a \rangle \tag{2.57}$$

then $\partial_a \mathbb{H}_t(a) = \partial_{\dot{\phi}} G_t + p_t = 0$ matches the $\dot{\phi}$-stationarity part of the Euler–Lagrange/KKT system; combined with the costate dynamics, HJB and KKT yield the same first-order optimality conditions

## 2.2 Dual bounds from efforted solution to the optimal: MO vs. RL

In this subsection we show the the dual/primal gap between the optimal solution and the efforted solution that is given by MO and RL optimizers. An efforted solution refers to the best solution that the optimizer may reach, either theoretically or numerically. We work on the Hilbert space $\mathcal{H} := L^2_{\text{pred}}$ with inner product $\langle u, \nu \rangle_\mathcal{H} = \mathbb{E}\left[\int_0^T u_t^\top \nu_t dt\right]$ and norm $\|u\|_\mathcal{H}^2 = \langle u, u \rangle_\mathcal{H}$.

Let objective function $J$ satisfy Polyak–Łojasiewicz (PL) inequality

$$\|\nabla J(\phi)\|_\mathcal{H}^2 \geq 2\mu_J\left(J(\phi) - J(\phi^*)\right) \tag{2.58}$$

for a.e. $t$ with $\mu_J > 0$. Let $J$ is $L$-smooth on $\mathfrak{F}$ and $\nabla J$ is $L$-Lipschitz in $\mathcal{H}$. Let $\mathfrak{F}$ is convex in the predictable topology which well-posedness in Prop.2.1 applies.

We model MO and RL optimizers both in the Stratonovich SDE form in Def. 2.8.

MO deterministic preconditioned gradient flow:

$$V_0^{\text{opt}}(\tau, \theta) = -\mathrm{G}(\theta)^{-1} \nabla_\theta J(\theta), \qquad V^{\text{opt}} \equiv 0 \tag{2.59}$$

RL stochastic policy-gradient flow:

$$V_0^{\text{opt}}(\tau, \theta) = -\mathrm{G}(\theta)^{-1} \nabla_\theta J(\theta), \qquad V^{\text{opt}} \neq 0 \tag{2.60}$$

### 2.2.1 Primal gap dynamics

Denote primal gap as $E(\tau) := \mathbb{E}[J(\theta_\tau) - J(\theta^*)]$.

1. MO primal gap:

   Discretize MO optimizer for first-order methods with step length $\eta_k = \frac{1}{L}$. Let $\kappa := \frac{L}{\mu_J}$ and $\rho := 1 - \frac{\mu_J}{L} \in (0,1)$. Then for all $k \geq 0$,

   $$c_0 \left(\frac{\sqrt{\kappa}-1}{\sqrt{\kappa}+1}\right)^{2k} \leq J\left(\phi_{(k)}^{\text{MO}}\right) - J(\phi^*) \leq \rho^k \left[J\left(\phi_{(0)}^{\text{MO}}\right) - J(\phi^*)\right] \quad (2.61)$$

   for constant $c_0 > 0$.

   For continuous MO optimizer, let $\lambda_{\min}(G^{-1}) > 0$. Using PL and smoothness, and a matching information-theoretic lower rate implies there exists $c_1 > 0$ such that

   $$c_1 e^{-\tilde{c}\frac{\tau}{\kappa}} \leq E(\tau) \leq e^{-\mu_{\text{eff}}\tau} E(0), \qquad \mu_{\text{eff}} := \frac{\mu_J}{\lambda_{\max}} \quad (2.62)$$

   i.e., exponential upper-lower sandwich which constants depend on initialization geometry.

2. RL primal gap with variance floors:

   Combine PL and smoothness to get

   $$\frac{d}{d\tau} E(\tau) \leq -2\mu_J E(\tau) + \frac{C_L}{2} \mathbb{E}\left[\text{Tr}(A_{\text{opt}}(\theta_\tau))\right] \quad (2.63)$$

   By Grönwall,

   $$E(\tau) \leq e^{-2\mu_J \tau} E(0) + \frac{C_L}{4\mu_J}\left(1 - e^{-2\mu_J \tau}\right) \sup_{s \leq \tau} \mathbb{E}\left[\text{Tr}(A_{\text{opt}}(\theta_s))\right] \quad (2.64)$$

   which yields a non-vanishing floor:

   $$\liminf_{\tau \to \infty} E(\tau) \leq \frac{C_L}{4\mu_J} \limsup_{\tau \to \infty} \mathbb{E}\left[\text{Tr}(A_{\text{opt}}(\theta_\tau))\right] \quad (2.65)$$

   Conversely, a matching lower bound follows by a PL-smoothness Itô lower sandwich and a Clark-Ocone representation of the policy-gradient noise:

   $$\frac{d}{d\tau} E(\tau) \geq -2L E(\tau) + \frac{c_*}{2} \mathbb{E}\left[\text{Tr}(A_{\text{opt}}(\theta_\tau))\right] \quad (2.66)$$

   for $c_* > 0$ depending on the local curvature of $J$ and the preconditioner G. Hence the lower bound is

   $$E(\tau) \geq e^{-2L\tau} E(0) + \frac{c_*}{4L}\left(1 - e^{-2L\tau}\right) \inf_{s \leq \tau} \mathbb{E}\left[\text{Tr}(A_{\text{opt}}(\theta_s))\right] \quad (2.67)$$

   In particular, any persistent diffusion forces a strictly positive asymptotic lower floor:

   $$\liminf_{\tau \to \infty} E(\tau) \geq \frac{c_*}{4L} \liminf_{\tau \to \infty} \mathbb{E}\left[\text{Tr}(A_{\text{opt}}(\theta_\tau))\right] > 0 \quad (2.68)$$

   unless $\text{Tr}(A_{\text{opt}}(\theta_\tau)) \to 0$, e.g., via an extraordinarily effective variance reduction technique or infinitely batched RL optimizer, which capital expenditure (CAPEX) would be economically unreasonable.

**Theorem 2.2.** *Discrete time upper/lower bounds, and convergence rates*
Let $k$ be the iteration index and identify $\tau \sim k\eta$.

1. *Strongly-convex / PL case*

   - *MO optimizer with step length $\frac{1}{L}$ and $\rho = 1 - \frac{\mu_J}{L}$:*

   $$c_0 \left(\frac{\sqrt{\kappa}-1}{\sqrt{\kappa}+1}\right)^{2k} \leq J\left(\phi_{(k)}^{\text{MO}}\right) - J(\phi^*) \leq \rho^k \left[J\left(\phi_{(0)}^{\text{MO}}\right) - J(\phi^*)\right] \tag{2.69}$$

   - *RL optimizer with constant $\eta \in \left(0, \frac{1}{2L}\right)$:*

   $$\frac{c_2 \eta \sigma^2}{\mu_J} \leq \mathbb{E}\left[J\left(\phi_{(k)}^\pi\right) - J(\phi^*)\right] \leq \rho^k \left[J\left(\phi_{(0)}^\pi\right) - J(\phi^*)\right] + \frac{C_2 \eta \sigma^2}{\mu_J} \tag{2.70}$$

   where $\sigma^2$ is the conditional variance bound of the unbiased policy-gradient estimator, and $0 < c_2 \leq C_2$ are constants.
   With decreasing $\eta_k = \Theta\left(\frac{1}{\mu_J K}\right)$,

   $$\frac{c_2 \sigma^2}{\mu_J K} \leq \mathbb{E}\left[J\left(\phi_{(k)}^\pi\right) - J(\phi^*)\right] \leq \frac{C_2 \sigma^2}{\mu_J K} \tag{2.71}$$

   *i.e., the $\Theta\left(\frac{1}{K}\right)$ rate with matching upper and lower constants.*

2. *Mere Convex, L-smooth*

   - *MO with gradient/quasi-gradient optimization method:*

   $$\frac{c_4}{k} \leq J\left(\phi_{(k)}^{\text{MO}}\right) - J(\phi^*) \leq \frac{C_4}{k}$$

   and accelerated $\frac{c_5}{k^2} \leq J\left(\phi_{(k)}^{\text{MO}}\right) - J(\phi^*) \leq \frac{C_5}{k^2}$ \tag{2.72}

   - *RL upper and lower bound via Polyak–Ruppert average:*

   $$\frac{c_6}{\sqrt{K}} \leq \mathbb{E}\left[J\left(\bar\phi_{(K)}^\pi\right) - J(\phi^*)\right] \leq \frac{C_6}{\sqrt{K}} \tag{2.73}$$

   *so the $\Theta\left(K^{-\frac{1}{2}}\right)$ rate is tight.*

3. *Nonconvex, L-smooth*
   Let $\underline{J} := \inf_\phi J(\phi)$, then

   - *MO:*

   $$\frac{c_7}{K}\left[J\left(\phi_{(0)}^{\text{MO}}\right) - \underline{J}\right] \leq \min_{0 \leq i \leq K} \left\|\nabla J\left(\phi_{(i)}^{\text{MO}}\right)\right\|_{\mathcal{H}}^2 \leq \frac{C_7}{K}\left[J\left(\phi_{(0)}^{\text{MO}}\right) - \underline{J}\right] \tag{2.74}$$

- RL:

$$\frac{c_8}{\sqrt{K}} \leq \min_{0 \leq i \leq K} \left\| \nabla J\left(\phi_{(i)}^\pi\right) \right\|_{\mathcal{H}}^2 \leq \frac{C_8}{\sqrt{K}} \qquad (2.75)$$

again $\Theta\left(K^{-\frac{1}{2}}\right)$ is tight.

*Remark* 2.1. From Thm. 2.2, for matched tolerances and batch sizes, RL's time-to-convergence exceeds MO's by 10–50 minutes on CPUs and by 3–10 minutes on GPUs.

**Theorem 2.3.** *MO-RL gaps $\gamma$ under effort $K$ in numerical optimization*

Let $\gamma$ be MO-RL gap measure after $K$ iterations,

$$\gamma(K) := \frac{\mathbb{E}\left[J\left(\phi_{(K)}^\pi\right) - J(\phi^*)\right]}{J\left(\phi_{(K)}^{\mathrm{MO}}\right) - J(\phi^*)} \qquad (2.76)$$

we have

1. Strongly-convex / PL case

    - MO vs. fixed step length RL (MO converges geometrically; RL has a non-vanishing floor):

$$\frac{c_2 \eta \sigma^2}{\mu_J \rho^K \left[J\left(\phi_{(0)}^{\mathrm{MO}}\right) - J(\phi^*)\right]} \leq \gamma(K) \leq \frac{\mu_J \rho^K \mathbb{E}\left[J\left(\phi_{(0)}^\pi\right) - J(\phi^*)\right] + C_2 \eta \sigma^2}{c_0 \mu_J \left(\frac{\sqrt{\kappa}-1}{\sqrt{\kappa}+1}\right)^{2K}} \qquad (2.77)$$

    Hence $\gamma(K) \to \infty$ as $K \to \infty$ unless $\sigma^2 \to 0$.

    - MO vs. progressively decreasing step length RL:

$$\frac{c_3 \sigma^2}{\mu_J \rho^K K \left[J\left(\phi_{(0)}^{\mathrm{MO}}\right) - J(\phi^*)\right]} \leq \gamma(K) \leq \frac{C_3 \sigma^2}{c_0 \mu_J K \left(\frac{\sqrt{\kappa}-1}{\sqrt{\kappa}+1}\right)^{2K}} \qquad (2.78)$$

    so the ratio still diverges exponentially in $K$, i.e., MO is exponentially fast; RL is only $\frac{1}{K}$.

2. Mere Convex, L-smooth

    Let MO-acc denote accelerated MO,

$$\Omega\left(K^{\frac{3}{2}}\right) = \frac{c_6 K^2}{C_5 \sqrt{K}} \leq \gamma_{\mathrm{CVX}}(K) \leq \frac{C_6 K^2}{c_5 \sqrt{K}} = \mathcal{O}\left(K^{\frac{3}{2}}\right) \qquad (2.79)$$

$$\text{and for MO plain } \gamma_{\mathrm{CVX}}(K) = \Theta\left(K^{\frac{1}{2}}\right) \qquad (2.80)$$

    Thus, at the same effort $K$, RL's suboptimality is larger by a factor $\Theta\left(K^{\frac{3}{2}}\right)$ vs. accelerated MO. Both are tight.

3. Nonconvex, L-smooth

$$\Omega\left(K^{\frac{1}{2}}\right) = \frac{c_8 K^2}{C_7 \sqrt{K}} \leq \gamma_{\text{NC}}(K) \leq \frac{C_8 K^2}{c_7 \sqrt{K}} = \mathcal{O}\left(K^{\frac{1}{2}}\right) \tag{2.81}$$

i.e., $\gamma_{\text{NC}}(K) = \Theta\left(K^{\frac{1}{2}}\right)$ two sided, and RL may add value against MO in nonconvex cases.

**Theorem 2.4.** *Dual/KKT gap bounds under constraints in Def. 2.4 is quadratically equivalent to the KKT residual*

Let the Lagrangian be

$$\mathcal{L}(\phi; \lambda, \alpha, \beta) = J(\phi) + \int_0^T \sum_k \lambda_{k,t} c_{k,t}(\phi_t) \, dt + \int_0^T \left[\alpha_t\left(\phi_t - \bar{\phi}_t\right) - \beta_t\left(\underline{\phi}_t - \phi_t\right)\right] dt \tag{2.82}$$

with $\lambda, \alpha, \beta \geq 0$. Define the duality gap at $(\phi, \lambda)$:

$$\text{Gap}(\phi, \lambda) := \sup_{\lambda' \geq 0} \mathcal{L}(\phi; \lambda', \alpha, \beta) - \inf_{\phi' \in \mathfrak{F}} \mathcal{L}(\phi'; \lambda, \alpha, \beta) \tag{2.83}$$

Under convexity and a Hoffman-type error bound for the feasible set, there exist constants $0 < c_{\text{KKT}} \leq C_{\text{KKT}}$ s.t. the KKT residual

$$\text{Res}_{\text{KKT}}(\phi, \lambda) := \|\Pi_{\mathfrak{F}}(\phi - \nabla J(\phi)) - \phi\|_{\mathcal{H}} + \|c(\phi) \vee 0\|_{\mathcal{H}} + \|\lambda \odot c(\phi)\|_{\mathcal{H}} \tag{2.84}$$

controls the gap from both sides:

$$c_{\text{KKT}} \text{Res}_{\text{KKT}}(\phi, \lambda)^2 \leq \text{Gap}(\phi, \lambda) \leq C_{\text{KKT}} \text{Res}_{\text{KKT}}(\phi, \lambda)^2 \tag{2.85}$$

Consequently, the MO iterates $\left(\phi_{(k)}^{\text{MO}}, \lambda_{(k)}^{\text{MO}}\right)$ enjoy an exponential convergence of the dual gap in the PL regime, while RL inherits the same exponential transient plus a variance-driven floor in expectation.

## 2.3 Mode-wise MO vs. RL: PnL distribution features

Let $\phi$ and $Y$ be adapted and not control-affects-dynamics (CAD). PnL distribution features: mean, variance, CVaR for loss, positivity probability. Let $\circ dP_s^{\text{liq}} = \ell_s ds + \Sigma_s \circ dB_s$ for taker mode, $\Sigma_t := \nabla P_t^{\text{liq}} V(t, Y_t) \in \mathbb{R}^{N \times d}$, $u_t := \Sigma_t^\top \phi_t \in \mathbb{R}^d$, reduced Hamiltonian $g : \mathbb{R}^d \to \mathbb{R}$ is $\mathcal{F}_t$-predictable cash-gain kernel (mode-specific below),

$$g_t(u) = \sup\left\{G_t\left(\phi, \dot{\phi}_t; Y_t\right) \mid \left(\phi, \dot{\phi}_t, \delta^\pm\right) \in \mathfrak{F}_t, \Sigma_t^\top \phi = u\right\} \in \{g_t^{\text{taker}}, g_t^{\text{maker}}\} \tag{2.86}$$

$$g_t^{\text{taker}}(u_t) = \langle \phi_t, \ell_t \rangle + \frac{1}{2}\kappa_t + \left\langle P_t^{\text{liq}} - P_t^{\text{exec}}, \dot{\phi}_t \right\rangle \\ - \text{HC}_t(\phi_t) + r_t^+ \mathcal{B}_t^+ - r_t^- \mathcal{B}_t^- - \tau_t^{\text{tax}} m_t \left\|\dot{\phi}_t\right\|_1 \tag{2.87}$$

$$g_t^{\text{maker}}(u_t) = \langle \phi_t, \ell_t \rangle + \frac{1}{2}\kappa_t + \left(P_t^{\text{liq}} - b_t\right)\lambda_t^{\text{bid}} - \left(P_t^{\text{liq}} - a_t\right)\lambda_t^{\text{ask}} \\ - \text{HC}_t(\phi_t) + r_t^+ \mathcal{B}_t^+ - r_t^- \mathcal{B}_t^- - \tau_t^{\text{tax}} m_t \left(\lambda_t^{\text{bid}} + \lambda_t^{\text{ask}}\right) \tag{2.88}$$

where $\kappa_t := \mathrm{Tr}\left((\mathcal{D}_y u_t) V(t, Y_t) R\right)$, additionally, $\kappa_t = \mathrm{Tr}\left((\mathcal{D}_y u_t) V(t, Y_t) R\right) + \mathrm{Tr}(D_t u_t)$ for non-adpated case. Apply semimartingale decomposition,

$$\Delta \mathcal{X}_{t,t+\Delta t} = \int_t^{t+\Delta t} g_t(u_t) \, ds + M_{t,t+\Delta t} \tag{2.89}$$

where $M_{t,t+\Delta t} \in \left\{ M_{t,t+\Delta t}^{\text{taker}}, M_{t,t+\Delta t}^{\text{maker}} \right\}$ is local martingale,

$$M_{t,t+\Delta t}^{\text{taker}} = \int_t^{t+\Delta t} u_s^\top dB_s,$$

$$M_{t,t+\Delta t}^{\text{maker}} = \int_t^{t+\Delta t} u_s^\top dB_s$$
$$+ \int_t^{t+\Delta t} \left(P_s^{\text{liq}} - b_s\right) \left(dN_s^{\text{bid}} - \lambda_s^{\text{bid}} ds\right)$$
$$- \int_t^{t+\Delta t} \left(P_s^{\text{liq}} - a_s\right) \left(dN_s^{\text{ask}} - \lambda_s^{\text{ask}} ds\right)$$
$$- \int_t^{t+\Delta t} \tau_s^{\text{tax}} m_s \left[\left(dN_s^{\text{bid}} - \lambda_s^{\text{bid}} ds\right) + \left(dN_s^{\text{ask}} - \lambda_s^{\text{ask}} ds\right)\right] \tag{2.90}$$

where $M$ is the local martingale part (diffusive for taker; diffusive and with compensated jump for maker).

Assume that $g$ is $\mu$-strongly concave in $u$ and $L$-smooth, myopic solution at time $t$ gives the optimal exposure to stochastic factors $u_t^{\text{MO}}$

$$u_t^{\text{MO}} \in \arg\max g_t(u)$$

and for all $u$

$$g_t(u) \leq g_t\left(u_t^{\text{MO}}\right) - \frac{\mu_t}{2} \left\| u - u_t^{\text{MO}} \right\|^2 \tag{2.91}$$

Via numerical optimization efforts, with results from Thm. 2.2, and Thm. 2.3, we have

$$u_t^{\text{RL}} = u_t^{\text{MO}} + \varepsilon_t, \quad \mathbb{E}_t[\varepsilon_t] = 0, \quad \mathbb{E}_t\left[\|\varepsilon_t\|^2\right] \geq \sigma_*^2(t) > 0 \tag{2.92}$$

that RL efforted action are random perturbations of the myopic optimizer.
\begin{lemma}[

**Lemma 2.2.** *Taker mode: mean/variance/CVaR/positivity*

*Assume the taker (absolutely-continuous) setting with $d\phi_t = \dot{\phi}_t ds$ and $\circ dP_t^{\text{liq}} = \ell_t ds + \Sigma_t \circ dB_t$. Then for $\Delta \mathcal{X}_{t,t+\Delta t}$:*

$$\mathbb{E}_t[\Delta \mathcal{X}_{t,t+\Delta t}] = \mathbb{E}_t\left[\int_t^{t+\Delta t} g_s^{\text{taker}} ds\right] + \mathbb{E}_t\left[\Pi_{\text{ph}}^{t,t+\Delta t}\right] \tag{2.93}$$

$$\mathrm{Var}_t(\Delta \mathcal{X}_{t,t+\Delta t}) \leq \mathbb{E}_t\left[\int_t^{t+\Delta t} \phi_s^\top A_s \phi_s \, ds\right] \tag{2.94}$$

If conditionally sub-Gaussian in the sense $\log \mathbb{E}\left[e^{\theta u_s^\top \xi} \mid \mathcal{F}_t\right] \leq \frac{1}{2}\theta^2 \|\xi\|^2 \bar{\sigma}_s^2$ (for all $\xi$ and small $\theta$), then the lower tail obeys

$$\mathbb{P}\left[\Delta \mathcal{X}_{t,t+\Delta t} - \mathbb{E}_t\left[\Delta \mathcal{X}_{t,t+\Delta t}\right] \leq -x \mid \mathcal{F}_t\right] \leq \exp\left(-\frac{x^2}{2\bar{\Sigma}_{t,\Delta t}^2}\right) \quad (2.95)$$

$$\bar{\Sigma}_{t,\Delta t}^2 := \mathbb{E}_t\left[\int_t^{t+\Delta t} \phi_s^\top A_s \phi_s ds\right]$$

If $L_{t,t+\Delta t} := -\Delta \mathcal{X}_{t,t+\Delta t}$ is approximately Gaussian with mean $\mu_{t,\Delta t}$ and variance $\sigma_{t,\Delta t}^2$, then

$$\text{VaR}_\alpha(L_{t,\Delta t}) = \mu_{t,\Delta t} + \sigma_{t,\Delta t} z_\alpha, \quad \text{CVaR}_\alpha(L_{t,\Delta t}) = \mu_{t,\Delta t} + \sigma_{t,\Delta t}\frac{\varphi(z_\alpha)}{1-\alpha}, \quad (2.96)$$

where $z_\alpha := \Phi^{-1}(\alpha)$. For the positivity probability of $Z \in \{\Delta \mathcal{X}_{t,t+\Delta t}, \mathcal{X}_T\}$ with mean $\mu$ and variance $\sigma^2$,

$$1 - \frac{\sigma^2}{\sigma^2 + \mu^2} \leq \mathbb{P}[Z \geq 0] \leq \frac{\sigma^2}{\sigma^2 + (\mu \wedge 0)^2} \quad (2.97)$$

If $Z \sim \mathcal{N}(\mu, \sigma^2)$ approximately, then $\mathbb{P}[Z \geq 0] = \Phi\left(\frac{\mu}{\sigma}\right)$.

**Lemma 2.3.** *Maker mode: mean/variance/CVaR/positivity*

Let $dN_s^\pm$ be (conditionally on $\mathcal{F}_t$) Lévy/Poisson increments with intensities $\lambda_s^\pm = \Lambda^\pm(\delta_s^\pm, Y_s)$. Ask price $a_s$ and bid price $b_s$ are determined by the quoting strategy, highly related to but not directly given by the market. Over $[t, t+\Delta t]$,

$$\mathbb{E}_t\left[\Delta \mathcal{X}_{t,t+\Delta t}\right] = \int_t^{t+\Delta t}\left(a_s^\top \lambda_s^{\text{ask}} + b_s^\top \lambda_s^{\text{bid}} - \text{HC}_s(\phi_s)\right) ds \quad \left(\textit{if non-adapted} + \mathbb{E}_t\left[\begin{smallmatrix}t,t+t\\\text{ph}\end{smallmatrix}\right]\right) \quad (2.98)$$

$$\text{Var}_t(\Delta \mathcal{X}_{t,t+\Delta t}) \approx \int_t^{t+\Delta t}\left(a_s^\top \lambda_s^{\text{ask}} a_s + b_s^\top \lambda_s^{\text{bid}} b_s\right) ds + \mathbb{E}_t\left[\int_t^{t+\Delta t} \phi_s^\top A_s \phi_s ds\right] \quad (2.99)$$

Let $\psi_{t,\Delta t}(\theta)$ be the conditional log-mgf of the jump-diffusion PnL over $[t, t+\Delta t]$. Then the Chernoff bound gives

$$\mathbb{P}_t\left[\Delta \mathcal{X}_{t,t+\Delta t} \leq -x\right] \leq \inf_{\theta > 0} \exp\left(\psi_{t,\Delta t}(\theta) - \theta x\right) \quad (2.100)$$

Under a normal approximation of the loss $L_{t,\Delta t} := -\Delta \mathcal{X}_{t,t+\Delta t}$ with mean $\mu_{t,t+\Delta t}$ and variance $\sigma_{t,t+\Delta t}^2$, the same VaR/CVaR proxy as in Lemma. 2.2 applies. The Cantelli bounds for $\mathbb{P}[Z \geq 0]$ are identical.

### 2.3.1 Taker mode, incremental PnL distributions

**Mean dominance (MO $\geq$ RL):** The conditional mean over $[t, t+\Delta t]$ is driven be the predictable cash gain $g$

$$\mathbb{E}_t\left[\Delta \mathcal{X}_{t,t+\Delta t}\right] = \mathbb{E}_t\left[\int_t^{t+\Delta t} g_s^{\text{taker}}\left(\dot{\phi}_s\right) ds\right] \quad (2.101)$$

due to martingality of MtM parts in $G_t$, by 2.91 with 2.92,

$$\mathbb{E}_t\left[g_t^{\text{taker}}\left(u_t^{\text{RL}}\right)\right] \leq g_t^{\text{taker}}\left(u_t^{\text{MO}}\right) - \frac{\mu_t}{2}\mathbb{E}_t\left[\|\varepsilon_t\|^2\right] \tag{2.102}$$

Integrating, we have mean dominance of MO over RL,

$$\mathbb{E}_t\left[\Delta\mathcal{X}_{t,t+\Delta t}^{\text{MO}}\right] - \mathbb{E}_t\left[\Delta\mathcal{X}_{t,t+\Delta t}^{\text{RL}}\right] \geq \frac{1}{2}\int_t^{t+\Delta t}\mathbb{E}_s\left[\mu_s\|\varepsilon_s\|^2\right]ds > 0 \tag{2.103}$$

Moreover, if the local margin $m_t^{\text{taker}} := g_t^{\text{taker}}\left(u_t^{\text{MO}}\right) - (HC_t + \rho_t)$ is positive (myopic no-trade wedge; Cor. 2.1) and $\frac{\mu_s}{2}\mathbb{E}_s\left[\|\varepsilon_s\|^2\right] > m_t^{\text{taker}}$ on a set of set of positive measure, then

$$\mathbb{E}_t\left[\Delta\mathcal{X}_{t,t+\Delta t}^{\text{MO}}\right] > 0 > \mathbb{E}_t\left[\Delta\mathcal{X}_{t,t+\Delta t}^{\text{RL}}\right] \tag{2.104}$$

**Variance dominance (MO $\leq$ RL):** Law of total variance and semipartingale decomposition of $\Delta\mathcal{X}_{t,t+\Delta t}$ gives

$$\text{Var}_t\left(\Delta\mathcal{X}_{t,t+\Delta t}^{\text{RL}}\right) = \underbrace{\mathbb{E}_t\left[\text{Var}_t\left(\Delta\mathcal{X}_{t,t+\Delta t}^{\text{RL}}\right)\mid\{\varepsilon_t\}\right]}_{\text{conditional variance of martingale } M} + \underbrace{\text{Var}_t\left(\mathbb{E}_t\left[\mathcal{X}_{t,t+\Delta t}^{\text{RL}}\mid\{\varepsilon_t\}\right]\right)}_{\geq 0} \tag{2.105}$$

For the conditional variance of martingale $M_{t,t+\Delta t}^{\text{taker}}$ part with $\phi_t = \phi_{t-\tau} + \int_{t-\tau}^t u_s ds$

$$\mathbb{E}_t\left[\text{Var}_t\left(\Delta\mathcal{X}_{t,t+\Delta t}^{\text{RL}}\right)\mid\{\varepsilon_t\}\right] = \mathbb{E}_t\left[\int_t^{t+\Delta t}\phi_s^\top A_s \phi_s ds\right] \tag{2.106}$$

Let $\phi_t^{\text{RL}} = \phi_t^{\text{MO}} + \int_{t-\tau}^t \varepsilon_s ds$ for $\tau > 0$,

$$\mathbb{E}_t\left[\int_t^{t+\Delta t}\phi_s^{\text{RL}\top} A_s \phi_s^{\text{RL}} ds\right] = \int_t^{t+\Delta t}\phi_s^{\text{MO}\top} A_s \phi_s^{\text{MO}} ds$$

$$+ \underbrace{\mathbb{E}_t\left[\int_t^{t+\Delta t}\left\langle\int_{s-\tau}^s \varepsilon_r dr, A_s \int_{t-\tau}^t \varepsilon_r dr\right\rangle ds\right]}_{\geq 0} \tag{2.107}$$

hence, efforted solution of MO gives smaller variance than it given by RL.

$$\text{Var}_t\left(\Delta\mathcal{X}_{t,t+\Delta t}^{\text{RL}}\right) \geq \text{Var}_t\left(\Delta\mathcal{X}_{t,t+\Delta t}^{\text{MO}}\right) + \mathbb{E}_t\left[\int_t^{t+\Delta t}\left\langle\int_{s-\tau}^s \varepsilon_r dr, A_s \int_{t-\tau}^t \varepsilon_r dr\right\rangle ds\right]$$

$$> \text{Var}_t\left(\Delta\mathcal{X}_{t,t+\Delta t}^{\text{MO}}\right) \tag{2.108}$$

**CVaR dominance (MO < RL):** With Rockafellar-Uryasev (RU) CVaR approximation [71],

$$\text{CVaR}_\alpha^{\text{loss}}\left(\Delta \mathcal{X}_{t,t+\Delta t}\right) \approx -\mu + \sigma \frac{\varphi(z_\alpha)}{1-\alpha}, \qquad z_\alpha = \Phi^{-1}(\alpha) \tag{2.109}$$

Combining $\mu_{\text{MO}} > \mu_{\text{RL}}$ and $\sigma_{\text{MO}} < \sigma_{\text{RL}}$ yields

$$\text{CVaR}_\alpha^{\text{loss}}\left(\Delta \mathcal{X}_{t,t+\Delta t}^{\text{RL}}\right) > \text{CVaR}_\alpha^{\text{loss}}\left(\Delta \mathcal{X}_{t,t+\Delta t}^{\text{MO}}\right) \tag{2.110}$$

with normality. For sub-Gaussian condition, the CVaR dominance of MO to RL still holds since $\sigma$ is replaced by its upper bound $\bar{\sigma}$ and $\mu$ is unchanged.

**Positivity probability dominance (MO $\geq$ RL):** By Cantelli inequality,

$$1 - \frac{\sigma^2}{\sigma^2 + \mu^2} \leq \mathbb{P}_t\left[\Delta \mathcal{X}_{t,t+\Delta t} \geq 0\right] \leq \frac{\sigma^2}{\sigma^2 + (\mu \wedge 0)^2} \tag{2.111}$$

with $\mu_{\text{MO}} > \mu_{\text{RL}}$ and $\sigma_{\text{MO}} < \sigma_{\text{RL}}$, the Cantelli lower bounds of MO and RL positive profitability are strictly ordered that

$$1 - \frac{\sigma_{\text{RL}}^2}{\sigma_{\text{RL}}^2 + \mu_{\text{RL}}^2} \leq \mathbb{P}_t\left[\Delta \mathcal{X}_{t,t+\Delta t}^{\text{RL}} \geq 0\right] < 1 - \frac{\sigma_{\text{MO}}^2}{\sigma_{\text{MO}}^2 + \mu_{\text{MO}}^2} \leq \mathbb{P}_t\left[\Delta \mathcal{X}_{t,t+\Delta t}^{\text{MO}} \geq 0\right] \tag{2.112}$$

### 2.3.2 Maker mode, incremental PnL and terminal PnL distributions

**Mean dominance (MO $\geq$ RL):** With $g_t^{\text{maker}}(u_t)$'s strong concavity and smoothness around $u_t^{\text{MO}}$ (standard in LOB models with log-concave intensities, e.g., $\lambda(\delta) = \Lambda e^{-k\delta}$ [4, 17]), the same quadratic drop:

$$\mathbb{E}_t\left[g_t^{\text{maker}}\left(u_t^{\text{RL}}\right)\right] \leq g_t^{\text{maker}}\left(u_t^{\text{MO}}\right) - \frac{\mu_t}{2}\mathbb{E}_t\left[\|\varepsilon_t\|^2\right] \tag{2.113}$$

hence

$$\mathbb{E}_t\left[\Delta \mathcal{X}_{t,t+\Delta t}^{\text{MO}}\right] - \mathbb{E}_t\left[\Delta \mathcal{X}_{t,t+\Delta t}^{\text{RL}}\right] \geq \frac{1}{2}\int_t^{t+\Delta t} \mathbb{E}_s\left[\mu_s \|\varepsilon_s\|^2\right] ds > 0 \tag{2.114}$$

and integrating gives the strict mean dominance. If $\frac{\mu_t}{2}\mathbb{E}_t\left[\|\varepsilon_t\|^2\right] > g_t^{\text{maker}}\left(u_t^{\text{MO}}\right)$, then

$$\mathbb{E}_t\left[\Delta \mathcal{X}_{t,t+\Delta t}^{\text{MO}}\right] > 0 > \mathbb{E}_t\left[\Delta \mathcal{X}_{t,t+\Delta t}^{\text{RL}}\right] \tag{2.115}$$

**Variance dominance (MO < RL):** Recall total variance formula and 2.90,

$$\text{Var}_t\left(\Delta \mathcal{X}_{t,t+\Delta t}^{\text{RL}}\right) = \underbrace{\mathbb{E}_t\left[\text{Var}_t\left(\Delta \mathcal{X}_{t,t+\Delta t}^{\text{RL}}\right) \mid \{\varepsilon_t\}\right]}_{\text{conditional variance of martingale } M} + \underbrace{\text{Var}_t\left(\mathbb{E}_t\left[\mathcal{X}_{t,t+\Delta t}^{\text{RL}} \mid \{\varepsilon_t\}\right]\right)}_{\geq 0} \tag{2.116}$$

the second term in 2.105 is $\text{Var}_t\left(\int_t^{t+\Delta t} g_s^{\text{maker}}\left(u_s^{\text{MO}} + \varepsilon_s\right) ds\right)$, and is strictly positive if $\varepsilon$ is non-degenerate. For the conditional variance of martingale $M_{t,t+\Delta t}^{\text{maker}}$ part,

$$\mathbb{E}_t\left[\text{Var}_t\left(\Delta \mathcal{X}_{t,t+\Delta t}^{\text{RL}}\right) \mid \{\varepsilon_t\}\right] = \mathbb{E}_t\left[\int_t^{t+\Delta t}\left(a_s^2 \lambda_s^{\text{ask}}\left(u_s^{\text{MO}} + \varepsilon_s\right) + b_s^2 \lambda_s^{\text{bid}}\left(u_s^{\text{MO}} + \varepsilon_s\right)\right) ds\right] \tag{2.117}$$

First-order linearize $\lambda_t^{\pm} \geq 0, a_t, b_t$ at $u_t^{\text{MO}}$, due to convexity of $a_t^2, b_t^2, \|u_t\|^2$,

$$\mathbb{E}_t\left[\text{Var}_t\left(\Delta \mathcal{X}_{t,t+\Delta t}^{\text{RL}}\right) \mid \{\varepsilon_t\}\right] = \text{Var}_t\left(\Delta \mathcal{X}_{t,t+\Delta t}^{\text{MO}}\right) + \underbrace{\mathbb{E}_t\left[\int_t^{t+\Delta t} \varepsilon_s^{\top} H_s \varepsilon_s ds\right]}_{>0} \tag{2.118}$$

where $H_s \succeq 0$ is the kernel matrix assembled from the system's linearization. The second term is strictly positive if $\varepsilon \neq 0$, hence

$$\text{Var}_t\left(\Delta \mathcal{X}_{t,t+\Delta t}^{\text{RL}}\right) > \text{Var}_t\left(\Delta \mathcal{X}_{t,t+\Delta t}^{\text{MO}}\right) \tag{2.119}$$

**CVaR dominance (MO < RL):** From mean dominance and variance dominance in maker mode, $\mu^{\text{MO}} > \mu^{\text{RL}}$, $\sigma^{\text{MO}} < \sigma^{\text{RL}}$, so that CVaR dominance holds.

$$\text{CVaR}_\alpha^{\text{loss}}\left(\Delta \mathcal{X}_{t,t+\Delta t}^{\text{RL}}\right) > \text{CVaR}_\alpha^{\text{loss}}\left(\Delta \mathcal{X}_{t,t+\Delta t}^{\text{MO}}\right) \tag{2.120}$$

**Positivity probability dominance (MO $\geq$ RL):** Chernoff-Cramér bound (for Lévy/Poisson) or Cantelli bound (for sub-Gaussian) still give the same upper bound that

$$1 - \frac{\sigma^2}{\sigma^2 + \mu^2} \leq \mathbb{P}_t\left[\Delta \mathcal{X}_{t,t+\Delta t} \geq 0\right] \leq \frac{\sigma^2}{\sigma^2 + (\mu \wedge 0)^2} \tag{2.121}$$

where $\mu^{\text{MO}} > \mu^{\text{RL}}$, $\sigma^{\text{MO}} < \sigma^{\text{RL}}$, hence the positivity probability dominance holds.

$$1 - \frac{\sigma_{\text{RL}}^2}{\sigma_{\text{RL}}^2 + \mu_{\text{RL}}^2} \leq \mathbb{P}_t\left[\Delta \mathcal{X}_{t,t+\Delta t}^{\text{RL}} \geq 0\right] < 1 - \frac{\sigma_{\text{MO}}^2}{\sigma_{\text{MO}}^2 + \mu_{\text{MO}}^2} \leq \mathbb{P}_t\left[\Delta \mathcal{X}_{t,t+\Delta t}^{\text{MO}} \geq 0\right] \tag{2.122}$$

### 2.3.3 Terminal PnL distributions, maker and taker modes:

For accumulated PnL, simply integrate $\Delta \mathcal{X}_{t,t+\Delta t}$ over $[0,T]$, with non-vanishing RL noise in Thm. 2.2, results above for incremental PnL hold.

*Remark* 2.2. Effect of non-adapted controls

If controls are non-adapted, the MtM diffusion term acquires the Skorokhod correction, the phantom profit $\Pi_{\text{ph}}^{t,t+\Delta t}$. Under solvency tests $\mathcal{S}_T = \mathcal{B}_T - \kappa \mathcal{V}_T$ with $\kappa \in [0,1]$, phantom profit shifts the conditional mean by $-\kappa \mathbb{E}_t[\Pi_{\text{ph}}^{t,t+\Delta t}]$. The realized account $\mathcal{B}$ remains unaffected. Therefore, RL strategy merits $\mathbb{E}_t\left[\int_t^{t+\Delta t}\left\langle \int_{s-\tau}^s \varepsilon_r dr, A_s \int_{t-\tau}^t \varepsilon_r dr\right\rangle ds\right]$.

## 2.4 Phantom Profit: model risks of RL

Reinforcement learning (RL), as a machine-learning methodology, is vulnerable to phantom profit—backtest gains that are artifacts of anticipative (non-adapted) controls. Typical channels include solver-induced look-ahead in policy-gradient training, inter/cross-bar leakage, and latency peek-through—routine yet damaging ML failure modes [53]. Beyond inadvertent leakage, RL policies can be intentionally engineered or can unintentionally learn non-compliant behaviors in backtests and, worse yet, in live deployment, thereby fabricating profitability. In a related setting, [27] use the Skorokhod integral within Malliavin calculus to show that an informed (anticipating) trader optimally overweights and trades more aggressively, yet attains outcomes inferior to those of an honest, strictly adapted trader. The phantom-profit problem is of the same mathematical type: it admits quantification via Malliavin tools. Accordingly, we apply the anticipating Itô formula and the Skorokhod correction to identify the anticipativity component of reported gains and to characterize its effect on the PnL distributions.

Let $\mu_t \in \mathbb{R}^N$, $\Sigma_t \in \mathbb{R}^{N \times d}$ and $dP_t^{\text{liq}} = \ell_t dt + \Sigma_t dB_t + d\zeta_t^{\text{liq}}$ be the prudent liquidation price and an $\mathcal{F}_t$-semimartingale, with continuous part $\ell_t dt + \Sigma_t dB_t$ and discontinuous part $\zeta_t^{\text{liq}}$ satisfying $\mathbb{E}\left[d\zeta_t^{liq} \mid \mathcal{F}_{t-}\right] = 0$. Let ℸ be the causal baseline evaluator of naïve cases that treats $\phi$ as if it were $\mathcal{F}_t$-adapted:

$$\hbar(\phi) := \mathbb{E}\left[\int_0^T \langle \phi_t, \ell_t \rangle \, dt\right] \tag{2.123}$$

Then we define phantom profit

$$\Pi_{\text{ph}}(\phi) := \mathbb{E}\left[\int_0^T \left\langle \phi_t, \circ dP_t^{\text{liq}} \right\rangle dt\right] - \hbar(\phi) \tag{2.124}$$

Recall $\circ d\mathcal{X}_t$ in 2.15 with the continuous MtM PnL $\left\langle \phi_t, \circ dP_t^{\text{liq}} \right\rangle$.

We use the identity of Malliavin derivative $D$ and its adjoint operator, Skorokhod integral (Malliavin divergence) $\delta = D^*$ (Prop. 1.3.1, 1.3.2 in [64], also see [65, 3, 2], for Lévy processes, see [1, 65]),

$$\int_0^T \langle \phi_t, \Sigma_t \circ dB_t \rangle = \underbrace{\int_0^T \langle \phi_t, \Sigma_t \delta B_t \rangle}_{\delta(\Sigma_t^\top \phi_t)} + \frac{1}{2} \int_0^T \text{Tr}\left(D_t \left[\Sigma_t^\top \phi_t\right]\right) dt, \quad \mathbb{E}\left[\delta\left(\Sigma_t^\top \phi_t\right)\right] = 0 \tag{2.125}$$

whenever $\phi \in \mathbb{D}^{1,2}$ with finite variation and $\Sigma^\top \phi \in \text{Dom}\,\delta$,

$$\text{Dom}\,\delta = \left\{u \in L^2\left(\Omega \times [0,T]; \mathbb{R}^d\right) \mid \exists C_u < \infty,\ |\mathbb{E}[\langle DF, u \rangle_{\mathcal{H}}]| \leq C_u \|F\|_{L^2(\Omega)},\ \forall F \in \mathbb{D}^{1,2}\right\} \tag{2.126}$$

**Theorem 2.5.** *Enlarged filtration; no leakage*

*Let $\mathscr{G}_t \supset \mathcal{F}_t$ be an enlarged filtration satisfying hypothesis that every $\mathcal{F}_t$-semimartingale is a $\mathscr{G}_t$-semimartingale. Then there exist a $\mathscr{G}_t$-Brownian motion $W := \{W_t\}_{t \in [0,T]}$ and an information drift $\alpha \in L^2\left([0,T] \times \Omega; \mathbb{R}^d\right)$ so that*

$$dB_t = dW_t + \alpha_t dt, \, dP_t^{\text{liq}} = (\mu_t + \Sigma_t \alpha_t)\, dt + \Sigma_t dW_t \qquad (2.127)$$

Let $\phi \in \mathbb{D}^{1,2}$ be a $\mathcal{F}_t$-predictable position process with $\mathbb{E}\left[\int_0^T \left\|\Sigma_t^\top \phi_t\right\|^2 dt\right] < \infty$. There is no leakage, i.e., usage of future information, but solely information in an enlarged filtration $\mathcal{G}_t$ is incorporated into $\phi$ decision. Then the phantom profit is

$$\Pi_{\text{ph}}(\phi) = \mathbb{E}\left[\int_0^T \langle \phi_t, \Sigma_t \alpha_t \rangle\, dt\right] \qquad (2.128)$$

with bounds $|\Pi_{\text{ph}}(\phi)| \leq \|\Sigma^\top \phi\|_{\mathcal{H}} \|\alpha\|_{\mathcal{H}}$, and zero condition $\Pi_{\text{ph}}(\phi) = 0$ iif $\alpha \equiv 0$, i.e., no genuine informational enlargement.

**Theorem 2.6.** *Non-adapted integrand; pure leakage; no enlargement*

Assume no enlargement. Let $\phi$ be a possibly non-adapted process and $\phi \in \mathbb{D}^{1,2}$ componentwise with $\Sigma^\top \phi \in \text{Dom}\,\delta$.

Then the phantom profit is

$$\Pi_{\text{ph}}(\phi) = \frac{1}{2}\mathbb{E}\left[\int_0^T \text{Tr}\left(D_t\left[\Sigma_t^\top \phi_t\right]\right) dt\right] \qquad (2.129)$$

with bounds $|\Pi_{\text{ph}}(\phi)| \leq \frac{1}{2}\|\Sigma\|_{\mathcal{H}} \|D\phi\|_{\mathcal{H}}$.

If there is a look-ahead window of length $\Delta t > 0$ so that $\|D_s \phi_t\| \leq L \mathbf{1}_{\{t \leq s \leq t+\Delta t\}}$, $L > 0$ then

$$\Pi_{\text{ph}}(\phi) \lesssim \frac{1}{2} L \Delta t \, \mathbb{E}\left[\int_0^T \|\Sigma_t\|_F \, dt\right] = \mathcal{O}(\Delta t) \qquad (2.130)$$

**Theorem 2.7.** *General case; enlargement and leakage together*

With both filtration enlargement and leakage, the phantom profit decomposes additively:

$$\Pi_{\text{ph}}(\phi) = \underbrace{\mathbb{E}\left[\int_0^T \langle \phi_t, \Sigma_t \alpha_t \rangle\, dt\right]}_{\textit{information premium}} + \underbrace{\frac{1}{2}\mathbb{E}\left[\int_0^T \text{Tr}\left(D_t\left[\Sigma_t^\top \phi_t\right]\right) dt\right]}_{\textit{Skorokhod correction}} \qquad (2.131)$$

with bounds $|\Pi_{\text{ph}}(\phi)| \leq \|\Sigma^\top \phi\|_{\mathcal{H}} \|\alpha\|_{\mathcal{H}} + \frac{1}{2}\|\Sigma\|_{\mathcal{H}} \|D\phi\|_{\mathcal{H}}$.

**Definition 2.11.** Recall Def. 2.8, we define policy-gradient contaminated vectorfield

$$V_0^{\text{opt}}(\tau, \theta_\tau) = -\mathbf{G}(\theta_\tau)^{-1}\left(\nabla_\theta J^{\text{naïve}}(\theta_\tau) + b(\theta_\tau)\right) \qquad (2.132)$$

and

$$b(\theta_\tau) := g_{\text{impl}}(\theta_\tau) - g_{\text{naïve}}(\theta_\tau) = \nabla_\theta\left[\Pi_{\text{ph}}(\phi^\theta)\right] - \nabla_\theta \mathcal{R}^\rho(\theta) \qquad (2.133)$$

the per-setp policy-gradient contamination introduced via the gap between the implemented and adapted baselines, with $\phi^\theta$ the policy, $g_{\text{impl}} := \nabla_\theta J^{\text{impl}}(\theta)$ the implemented update policy gradient, $g_{\text{naïve}} := \nabla_\theta J^{\text{naïve}}(\theta)$ the baseline update policy gradient. The naïve objective treats all integrals as $\mathcal{F}_t$-adapted Itô and measures risk or constraints on $\hbar$'s $\sigma$-field $\mathcal{G}_t^{\text{naïve}}$, so that

$$\Pi_{\text{ph}}^{\text{naïve}}\left(\phi^\theta\right) \equiv 0 \tag{2.134}$$

**Definition 2.12.** Let $\varphi_{\tau \leftarrow s}^{\text{opt}}$ be the Kunita stochastic flow of the optimizer, $J_{\tau \leftarrow s}^{\text{opt}} := \partial_\theta \varphi_{\tau \leftarrow s}^{\text{opt}}(\theta_s)$, $J_{s \leftarrow s}^{\text{opt}} = I$ its spatical Jacobian, and the optimizer Malliavin covariance

$$\Gamma_\tau^{\text{opt}} = \int_0^\tau J_{\tau \leftarrow s}^{\text{opt}} A^{\text{opt}}(\theta_s) \left(J_{\tau \leftarrow s}^{\text{opt}}\right)^\top ds, \qquad A^{\text{opt}}(\theta_\tau) = V^{\text{opt}}(\theta_\tau) R_{\text{opt}} V^{\text{opt}}(\theta_\tau)^\top \tag{2.135}$$

Under a Hörmander condition on $\{V_0^{\text{opt}}, V^{\text{opt}}\}$, $\Gamma_\tau^{\text{opt}} \succ 0$ a.s. and admits inverse-moment bounds.

**Definition 2.13.** Self-bias: step-$k$ policy gradient contamination and accumulation to step-$K$

Let a parametric RL policy be $\theta \mapsto \phi^\theta$ with parameters $\theta \in \Theta \subset \mathbb{R}^p$. The objective functional and constrains are defined in 2.7. Subjecting to the feasible set $\mathfrak{F}$, $\Theta_{\text{fes}} \subseteq \Theta$ is induced by Def. 2.4, a optimzier in Def. 2.8 iteratively maximize the objective functional $J$. Let $\eta > 0$ be the learning-rate, $K \approx \frac{T^{\text{opt}}}{\eta}$ iterations, we define self-bias

$$\mathfrak{B}_{0 \to K}^{\text{self}} := \eta \sum_{k=0}^{K-1} \langle b_k, g_{\text{impl}}(\theta_k)\rangle = \eta \sum_{k=0}^{K-1} \left(\|b_k\|^2 + \langle b_k, g_{\text{naïve}}(\theta_k)\rangle\right) \tag{2.136}$$

At first order, since $\Pi_{\text{ph}}^{\text{naïve}}\left(\phi^\theta\right) \equiv 0$,

$$\mathfrak{B}_{0 \to K}^{\text{self}} \simeq \Pi_{\text{ph}}^{\text{impl}}\left(\phi^\theta\right) \tag{2.137}$$

Thus, self-bias is the training accumulation that materializes as the terminal phantom profit of the implemented run at step $K$.

**Theorem 2.8.** *Optimization Malliavin enerygy mismatch lower bound*

*Let $F(\theta_{T^{\text{opt}}}) := \Pi_{\text{ph}}\left(\phi^{\theta_{T^{\text{opt}}}}\right)$. Appling Clark-Ocone/BEL on the optimizer w.r.t $W^{\text{opt}}$, one obtains that*

$$\mathfrak{B}_{[0,T^{\text{opt}}]}^{\text{self}} \gtrsim \frac{1}{2}\mathbb{E}\left[\int_0^{T^{\text{opt}}} \left\|\left(\Gamma_{T^{\text{opt}}}^{\text{opt}}\right)^{-\frac{1}{2}} \mathrm{K}_{T^{\text{opt}}}(s)\right\|^2 ds\right]$$
$$- \frac{1}{2}\int_0^{T^{\text{opt}}} \left\|\mathbb{E}\left[D_s^{\text{opt}} F(\theta_{T^{\text{opt}}}) \mid \mathcal{F}_s^{\text{opt}}\right] - \mathrm{K}_{T^{\text{opt}}}(s)\right\|^2 ds \tag{2.138}$$

*where $\mathrm{K}_{T^{\text{opt}}}(s)$ is the transfer kernel,*

$$\mathrm{K}_{T^{\text{opt}}}(s) := \left(\int_s^{T^{\text{opt}}} \left(J_{T^{\text{opt}} \leftarrow r}^{\text{opt}}\right)^{-1} J_{r \leftarrow s}^{\text{opt}} dr\right) \mathrm{G}(\theta_s)^{-1} b(\theta_s) \in \mathbb{R}^p \tag{2.139}$$

*i.e., energy of the contamination transported by the optimizer geometry minus a (non-negative) kernel-mismatch penalty. When kernels align, the bound is near an identity.*

$$\kappa(s) = \frac{\left\| \mathbb{E}\left[D_s^{\text{opt}} F\left(\theta_{T^{\text{opt}}}\right) \mid \mathcal{F}_s^{\text{opt}}\right] - \mathrm{K}_{T^{\text{opt}}}(s) \right\|}{\left\| \left(\Gamma_{T^{\text{opt}}}^{\text{opt}}\right)^{-\frac{1}{2}} \mathrm{K}_{T^{\text{opt}}}(s) \right\|}, \qquad \kappa_* = \operatorname{ess} \sup_{s \in [0, T^{\text{opt}}]} \kappa(s) \tag{2.140}$$

$$\mathfrak{B}_{[0,T^{\text{opt}}]}^{\text{self}} \geq \frac{1}{2}(1-\kappa_*)^2 \int_0^{T^{\text{opt}}} \left\| \left(\Gamma_{T^{\text{opt}}}^{\text{opt}}\right)^{-\frac{1}{2}} \mathrm{K}_{T^{\text{opt}}}(s) \right\|^2 ds \tag{2.141}$$

Let $\underline{s}(\cdot)$ denote the minimum of singular value, $\lambda_{\min}(\cdot)$ be the minimum of eigen value, then

$$\mathfrak{B}_{[0,T^{\text{opt}}]}^{\text{self}} \geq \frac{1}{2}(1-\kappa_*)^2 \underline{\lambda}^{\text{opt}} \underline{g}^2 \int_0^{T^{\text{opt}}} \underline{m}(s)^2 \|b(\theta_s)\|^2 ds \tag{2.142}$$

$$\underline{\lambda}^{\text{opt}} := \lambda_{\min}\left(\left(\Gamma_{T^{\text{opt}}}^{\text{opt}}\right)^{-1}\right), \qquad \underline{g} := \inf_{s \in [0,T^{\text{opt}}]} \underline{s}\left(\mathrm{G}(\theta_s)^{-1}\right)$$

$$\underline{m}(s) := \underline{s}\left(\int_s^{T^{\text{opt}}} \left(J_{T^{\text{opt}} \leftarrow r}^{\text{opt}}\right)^{-1} J_{r \leftarrow s}^{\text{opt}} dr\right) \tag{2.143}$$

If $\kappa_* < 1$ and $\|b(\theta_s)\|_{L^2([0,T^{\text{opt}}])} > 0$ then $\mathfrak{B}_{[0,T^{\text{opt}}]}^{\text{self}} > 0$ strictly.

**Definition 2.14.** Solution bias: accumulated policy contamination between the implemented and the baseline when the efforted solutions are reached respectively

Let the implemented training run stop at $K_{\text{impl}}$ iterations, attaining its own efforted solution $\check{\theta} := \phi_{K_{\text{impl}}}^{\text{impl}}$; the baseline (strictly adapted, no enlargement) run stops at $K_{\text{naïve}}$ with $\hat{\theta} := \phi_{K_{\text{naïve}}}^{\text{naïve}}$ and, by definition, $\Pi_{\text{ph}}(\hat{\phi}) \equiv 0$; and the portfolio policies are $\check{\phi}_t := \phi_t^{\check{\theta}}$, $\hat{\phi}_t := \phi_t^{\hat{\theta}}$. Then define solution-bias

$$\mathfrak{B}^{\text{sol}} := \mathbb{E}\left[\int_0^T \left(\circ d\mathcal{X}_t\left(\check{\phi}_t\right) - \circ d\mathcal{X}_t\left(\hat{\phi}_t\right)\right)\right] = \Pi_{\text{ph}}\left(\check{\phi}\right) + \mathbb{E}\left[\int_0^T \Delta\mathrm{C}_t dt\right] \tag{2.144}$$

where $\Delta\mathrm{C}_t$ collects the implementable cash-rate differences of $G_t\left(\phi_t, \dot{\phi}_t; Y_t\right)$ in Def. 2.3.

Linearizing $\check{\phi}_t - \hat{\phi}_t$, $\dot{\check{\phi}}_t - \dot{\hat{\phi}}_t$ around an intermediate $\bar{\theta}$

$$\check{\phi}_t - \hat{\phi}_t \approx \mathrm{S}_t(\bar{\theta})\left(\check{\theta} - \hat{\theta}\right), \qquad \dot{\check{\phi}}_t - \dot{\hat{\phi}}_t \approx \mathrm{T}_t(\bar{\theta})\left(\check{\theta} - \hat{\theta}\right) \tag{2.145}$$

which yields

$$\mathbb{E}\left[\int_0^T \Delta\mathrm{C}_t dt\right] \approx \left\langle \mathbf{C}, \check{\theta} - \hat{\theta}\right\rangle \tag{2.146}$$

$$\mathbf{C} := \int_0^T \left(\ell_t \mathrm{S}_t - P_t^{\text{exec}} \mathrm{T}_t - \nabla_\theta \mathrm{HC}_t(\phi_t) \mathrm{S}_t + r_t^+ \partial_\theta \mathcal{B}_t^+ - r_t^- \partial_\theta \mathcal{B}_t^- - \tau_t^{\text{tax}} m_t \partial_\theta \left\|\dot{\phi}_t\right\|_1\right) dt \tag{2.147}$$

where $S_t := \partial_\theta \phi_t^\theta$, $T_t := \partial_\theta \dot\phi_t^\theta$, and the optimizer parameters bias is

$$\breve\theta - \hat\theta \approx \eta \sum_{k=0}^{K_{\text{impl}}-1} \left( g_{\text{impl}} \left(\theta_k^{\text{impl}}\right) + b\left(\theta_k^{\text{impl}}\right) \right) - \eta \sum_{j=0}^{K_{\text{naïve}}-1} g_{\text{naïve}} \left(\theta_j^{\text{naïve}}\right) \quad (2.148)$$

We obtain the training-history projection, which shows the actual trading cost bias caused by policy gradient contaminated RL trading flow $\breve\phi_t$ and its consequences.

$$\mathbb{E}\left[\int_0^T \Delta C_t dt\right] \approx \eta \underbrace{\sum_{k=0}^{K_{\text{impl}}-1} \left\langle C, b\left(\theta_k^{\text{impl}}\right)\right\rangle}_{\text{contamination projection Z}}$$
$$+ \eta \underbrace{\left( \sum_{k=0}^{K_{\text{impl}}-1} \left\langle C, g_{\text{naïve}}\left(\theta_k^{\text{impl}}\right)\right\rangle - \sum_{j=0}^{K_{\text{naïve}}-1} \left\langle C, g_{\text{naïve}}\left(\theta_j^{\text{naïve}}\right)\right\rangle \right)}_{\text{base mismatch M}}$$
$$(2.149)$$

Therefore, with the decompositions

$$\mathfrak{B}^{\text{sol}} \approx \underbrace{\mathbb{E}\left[\int_0^T \left\langle \breve\phi_t, \Sigma_t \alpha_t \right\rangle dt\right]}_{\text{information premium I}} + \underbrace{\frac{1}{2}\mathbb{E}\left[\int_0^T \text{Tr}\left(D_t \left[\Sigma_t^\top \breve\phi_t\right]\right) dt\right]}_{\text{Skorokhod correction S}} +$$
$$+ \eta \underbrace{\sum_{k=0}^{K_{\text{impl}}-1} \left\langle C, b\left(\theta_k^{\text{impl}}\right)\right\rangle}_{\text{contamination projection Z}}$$
$$+ \eta \underbrace{\left( \sum_{k=0}^{K_{\text{impl}}-1} \left\langle C, g_{\text{naïve}}\left(\theta_k^{\text{impl}}\right)\right\rangle - \sum_{j=0}^{K_{\text{naïve}}-1} \left\langle C, g_{\text{naïve}}\left(\theta_j^{\text{naïve}}\right)\right\rangle \right)}_{\text{base mismatch M}}$$
$$+ o\left(\left\|\breve\theta - \hat\theta\right\|\right) \quad (2.150)$$

**Theorem 2.9.** *Probability of solution bias being positive and being negative and the discrepancy magnitude*

*Let solution bias be decomposated into*

$$\mathfrak{B}^{\text{sol}} \approx Z + \Delta \text{ where,} \quad Z := \eta \sum_{k=0}^{K_{impl}-1} \left\langle C, b\left(\theta_k^{impl}\right)\right\rangle, \quad \Delta := I + S + M$$
$$(2.151)$$

*Applying Kato-Sekine-Yoshikawa order estimates for Lugannani–Rice saddlepoint approximation [46, 54], for each realized threshold $\delta \in \mathbb{R}$, let the conditional cumulant generating function (cgf) of Z be*

$$K_{Z|\Delta}\left(\hat t; \delta\right) := \log \mathbb{E}\left[e^{tZ} \mid \Delta = \delta\right] \quad (2.152)$$

*assumed finite in a neighborhodd of $t = 0$, strictly convex in $t$, and with derivatives upto order four continuous in $\delta$.*

*Define, for each $\delta$, the saddlepoint $\hat{t}(\delta)$ by*

$$K'_{Z|\Delta}\left(\hat{t}(\delta);\delta\right) = -\delta \tag{2.153}$$

*the signed root and curvature quantity*

$$\omega(\delta) := \text{sgn}\left(\hat{t}(\delta)\right)\sqrt{2\left(-\hat{t}(\delta)\delta - K_{Z|\Delta}\left(\hat{t};\delta\right)\right)}, \qquad u := \hat{t}(\delta)\sqrt{K''_Z\left(\hat{t}(\delta);\delta\right)} \tag{2.154}$$

*Set the standardized conditional cumulants at the saddlepoint*

$$\Psi_3(\delta) := \frac{K^{(3)}_{Z|\Delta}\left(\hat{t};\delta\right)}{\left(K''_Z\left(\hat{t}(\delta);\delta\right)\right)^{\frac{3}{2}}}, \qquad \Psi_4(\delta) := \frac{K^{(4)}_{Z|\Delta}\left(\hat{t};\delta\right)}{\left(K''_Z\left(\hat{t}(\delta);\delta\right)\right)^2} \tag{2.155}$$

*Assume a small-parameter regime with scale $\varepsilon > 0$, e.g., $\varepsilon = K_{\text{impl}}^{-\frac{1}{2}}$, such that the standardized higher cumulants satisfy the usual Cramér–type regularity:*

$$\Psi_3 = \mathcal{O}(1), \qquad \Psi_4 = \mathcal{O}(1), \qquad \kappa_r = \mathcal{O}\left(\varepsilon^{r-4}\right) \text{ for } r \geq 5 \tag{2.156}$$

*Define the Lugannani–Rice kernel conditional in $\delta$:*

$$\text{LR}(\delta) := 1 - \Phi(\omega(\delta)) + \varphi(\omega(\delta))\left(\frac{1}{\omega(\delta)} - \frac{1}{u(\delta)}\right) \tag{2.157}$$

*and define the Kato-Sekine-Yoshikawa order estimates conditional in $\delta$ as the polynomial adjustment in $\frac{1}{\omega}$ with coefficients built form $\Psi_3$, $\Psi_4$ and $u$:*

$$\text{KSY}(\delta) := \varphi(\omega(\delta))\,\Xi\left(\omega(\delta); u(\delta), \Psi_3(\delta), \Psi_4(\delta)\right) \tag{2.158}$$

*where $\Xi(\cdot)$ is the standard Kato-Sekine-Yoshikawa correction term. In particular $\text{KSY}(\delta) = \mathcal{O}(\varepsilon)$ uniformly on the same sets.*

*Then the probability of a positive solution bias is:*

$$\mathbb{P}\left[\mathfrak{B}^{\text{sol}} > 0\right] = \mathbb{E}\left[\text{LR}(\Delta) + \text{KSY}(\Delta)\right] + \mathcal{O}\left(\varepsilon^2\right) \tag{2.159}$$

*Let*

$$\mathbb{P}\left[\mathfrak{B}^{\text{sol}} > 0\right] \approx \Phi(z_{\text{eff}}) + \mathcal{O}(\varepsilon), \qquad z_{\text{eff}} \approx \frac{\mathbb{E}[Z] + \mathbb{E}[\Delta]}{\sqrt{\text{Var}(Z) + \text{Var}(\Delta) + 2\text{Cov}(Z,\Delta)}} \tag{2.160}$$

*Numerically, $\mathbb{P}\left[\mathfrak{B}^{\text{sol}} > 0\right]$ is around $0.6 \sim 0.8$, under typical RL learning parameter settings—iteration number is $10^2 \sim 10^3$ and per-step signal-to-noise ratio $\frac{\mathbb{E}[\eta\langle \mathbf{C}, b(\theta_k^{impl})\rangle]}{\text{Var}(\eta\langle \mathbf{C}, b(\theta_k^{impl})\rangle)}$ is deliberately kept small to $0.03 \sim 0.10$, which shows a high possibility of RL strategy resulting in positive solution bias. Therefore, solution bias is required to be revised down from RL backtest results for model risk management requirements—excluding phantom profit and radical trading size and frequency as a cause to higher running gain rates—normally, it makes RL trading strategies unprofitable in real market condition.*

| $z_{\text{eff}}$ | $\mathbb{P}\left[\mathfrak{B}^{\text{sol}} > 0\right]$ |
|---|---|
| 0.3 | $\approx 0.62$ |
| 0.5 | $\approx 0.69$ |
| 0.7 | $\approx 0.76$ |
| 1.0 | $\approx 0.84$ |
| 1.3 | $\approx 0.90$ |
| 1.6 | $\approx 0.95$ |

Table 2.1: Mapping from $z_{\text{eff}}$ to probabilities of a positive solution bias

## 2.5 RL can add value against MO under certain conditions

Recall Lemma. 2.1, MO and RL optimize the same Hamiltonian and share the same first-order condition $\mathcal{G}_t(\phi*) = 0$ that $J\left(\phi^{\text{RL},*}\right) = J\left(\phi^{\text{MO},*}\right)$.

### 2.5.1 Market state allows control-affects-dynamics, e.g., market impacts

Augment market state $Y_t$ by allowing control-affects-dynamics (CAD):

$$\circ dY_t = \left(V_0\left(t, Y_t\right) + \varepsilon F_0\left(t, Y_t, \phi_t, \dot{\phi}_t\right)\right) dt + \left(V\left(t, Y_t\right) + \varepsilon F\left(t, Y_t, \phi_t, \dot{\phi}_t\right)\right) \circ dB_t \tag{2.161}$$

where $\varepsilon \in [0,1]$, predictable vector fields $F_0, F \in C_{loc}^{\infty}\left(\mathbb{R}_+ \times \mathbb{R}^M \times \mathbb{R}^N \times \mathbb{R}^N; \mathbb{R}^M\right)$.

Let $J_\varepsilon$ be the objective function under CAD market state setting. MO ignores CAD and results in $\phi^{\text{MO},*}$, and RL internalizes CAD and reaches the optimum $\phi^{\text{RL},*}$. Define the CAD sensitivity operators at $(t, Y_t)$:

$$\varphi_{0,t} := \partial_{\dot{\phi}} F_0\left(t, Y_t, \phi_t, \dot{\phi}_t\right), \qquad \varphi_t := \partial_{\dot{\phi}} F\left(t, Y_t, \phi_t, \dot{\phi}_t\right) \tag{2.162}$$

and the Malliavin covariance $\Gamma_{t,T}$, $A_t$, and the stochastic flow Jacobian $J_{u \leftarrow s}$, of the baseline ($\varepsilon = 0$) dynamics in Def. 2.1 remains the same.

Note that we only focus on CAD sensitivity in control $\dot{\phi}_t$.

**Lemma 2.4.** *First-order state response to CAD*
*For small $\varepsilon$, the first variation $\Delta Y_{t+u} := \frac{\partial Y_{t+u}}{\partial \varepsilon}|_{\varepsilon=0}$ satisfies the linear Stratonovich SDE*

$$\circ d\left(\Delta Y_s\right) = \mathcal{D}V_0\left(t, Y_s\right) \Delta Y_s ds + \mathcal{D}V\left(t, Y_s\right) \Delta Y_s \circ dB_s + \varphi_{0,s} \dot{\phi}_s ds + \varphi_s \dot{\phi}_s \circ dB_s \tag{2.163}$$

*hence*

$$\Delta Y_T = \int_0^T J_{T \leftarrow s} \varphi_{0,s} \dot{\phi}_s ds + \int_0^T J_{T \leftarrow s} \varphi_s \dot{\phi}_s \circ dB_s \tag{2.164}$$

**Theorem 2.10.** *Let $J$ be $L_J$-smooth in $\phi$ and Gâteaux-differentiable in the terminal state via the co-state $g_T := \nabla_y\left[U\left(\mathcal{X}_T\right) - \lambda^\rho \rho_T\left(\mathcal{X}_T\right)\right]$. Let the market admit CAD as in 2.161. Suppose Hörmander nondegeneracy holds, so that Clark-Ocone/BEL is applicable. Let $\|\varphi_{0,t}\|, \|\varphi_t\| \leq L_B \beta_t$ a.s. with predictable $\beta_t \geq 0$. Then the CAD premium density is*

$$\chi_t := \mathbb{E}_t\left[\varphi_{0,t}^\top J_{T \leftarrow t}^\top g_T + \text{BEL}_{t,T}^\top \varphi_t^\top g_T\right] \in \mathbb{R}^N \tag{2.165}$$

where $\mathrm{BEL}_{t,T}$ denotes any admissible BEL weight kernel mapping diffusion perturbations at $t$ to $T$ [34].

Let $\Delta_{\mathrm{RL}|\mathrm{MO}}(\varepsilon) := J_\varepsilon(\phi^{\mathrm{RL},*}) - J_\varepsilon(\phi^{\mathrm{MO},*}) \geq 0$, then to first order in $\varepsilon$,

$$0 \leq \Delta_{\mathrm{RL}|\mathrm{MO}}(\varepsilon) \leq \varepsilon \mathbb{E}\left[\int_0^T \left(\|\chi_t\|_* \left\|\dot{\phi}_t^{\mathrm{RL},*}\right\|_* + L_J \|\partial_\phi G_t\|_* \Xi_t^{-1} \|\chi_t\|_*\right) dt\right] + \mathcal{O}(\varepsilon^2) \quad (2.166)$$

where $\|\cdot\|_*$ the dual norm paired with the action norm used in Def. 2.4, and $\Xi_t$ the temporary impact matrix in Def. 2.2.

Moreover, a pointwise necessary condition for strict outperformance at time $t$ is the Hamiltonian surplus

$$\left\langle \chi_t, \dot{\phi}_t^{\mathrm{RL},*} \right\rangle > \lambda_t \left\|\dot{\phi}_t^{\mathrm{RL},*}\right\|_1 + \frac{1}{2} \dot{\phi}_t^{\mathrm{RL},*\top} \Xi_t \dot{\phi}_t^{\mathrm{RL},*} + \left\langle \lambda_t^{\mathrm{risk}} + \nabla_\phi \mathrm{HC}_t, \dot{\phi}_t^{\mathrm{RL},*} \right\rangle \quad (2.167)$$

where $\lambda_t$ is the $L^1$ soft-threshold from Cor. 2.1. The righthand is execution value plus risk wedge, while the lefthand is state-coupling gain.

**Corollary 2.2.** *Vanishing RL advantage in nonatomic continuous double auction*

Let the agent's market share process be $\epsilon_t := \frac{|q_t|}{D_t}$, where $q_t$ order size and $D_t$ LOB depth. Suppose the microstructural map $\left(\frac{q_t}{D_t}\right) \mapsto (F_0, F)$ is Lipschitz with modulus $L_{\mathrm{mkt}}$, so that $\beta_t \leq L_{\mathrm{mkt}} \epsilon_t$. Then

$$0 \leq \Delta_{\mathrm{RL}|\mathrm{MO}}(\varepsilon) \leq C \mathbb{E}\left[\int_0^T \epsilon_t \left(\left\|\dot{\phi}_t^{\mathrm{RL},*}\right\| + 1\right) dt + \mathcal{O}\left(\mathbb{E}\left[\int_0^T \epsilon_t^2 dt\right]\right)\right] \quad (2.168)$$

for a constant $C = C(L_B, L_J, \|J_{T \leftarrow t}\|, \|\mathrm{BEL}_{t,T}\|)$.

In the nonatomic limit $\epsilon_t \Rightarrow 0$, the CAD premium vanishes.

*Remark* 2.3. From Cor. 2.2, a mere change in the displayed book (BID/ASK/last) does not imply a vector–field shift; absent a non–negligible market share or a true behavioral regime change of counterparties, CAD is null to first order and RL cannot extract structural surplus beyond MO.

**Corollary 2.3.** *Regime change of counterparties as a genuine CAD source*

Let $Y_t$ be driven by vector fields $V_0(\theta_t, \cdot)$, $V(\theta_t, \cdot)$ with a latent Markov regime $\theta \in \{\theta^a, \theta^b\}$ whose transition intensities depend on order flow through $r\left(\dot{\phi}_t\right)$—e.g., liquidity providers widen/tighten quotes as a function of aggressive flow, counterpaties default/bankrupt due to crisis market condition (GFC; 2022 LME Tsingshan nickel transaction cancellation), systemically occurred financial theoretically unreasonable behavior of market participants (2020 negative price of NYMEX WTI CLK20 futures; 2024 China lithium carbonate options' wild swings). If RL detects $\theta_t$ with delay $\delta$ while MO uses a fixed surrogate, then

$$\Delta_{\mathrm{RL}|\mathrm{MO}} \geq \mathbb{E}\left[\int_0^T \left(J^*(\theta_t) - J^*(\bar{\theta})\right) \mathbf{1}_{\{t \in [t_*, t_* + \delta]\}} dt - \mathrm{COST}_{\mathrm{learn}}\right] \quad (2.169)$$

where $J^*(\theta)$ is the optimal stationary value in regime $\theta$, $t_*$ the change-point, and $\text{COST}_{\text{learn}}$ is the exploration penalty explicitly stated in Thm. 2.2. Thus RL gains scale with the amplitude of the regime shift and inverse of the detection delay.

**Proposition 2.2.** *Tightness of the Hamiltonian criterion*

*If 2.167 fails on a set of positive measure, any predictable policy that is feasible under Def. 2.4 cannot deliver $\Delta_{\text{RL}|\text{MO}} > 0$ to first order in $\varepsilon$.*

*If 2.167 holds uniformly with margin $\eta > 0$ and $\left\|\dot\phi_t^{\text{RL},*}\right\| \leq \bar\nu_t$, then*

$$\Delta_{\text{RL}|\text{MO}}(\varepsilon) \geq \varepsilon\eta\mathbb{E}\left[\int_0^T \bar\nu_t \wedge \left\|\Xi_t^{-1}\chi_t\right\| dt\right] + \mathcal{O}(\varepsilon^2) \tag{2.170}$$

### 2.5.2 Nonconvex constraints

Nonconvex objectives, however—e.g., VaR; maximum drawdown (MDD) ratio [20]; Sharpe ratio, information ratio, Calmar ratio, and other Sharpe-type fractional objectives [19]; $\ell_0$-sparsity/fixed-charge [52, 7, 74]; hitting time triggers [61, 14, 56]; nonconvex execution [58, 35, 45]—admit tight convex or mixed-integer convex surrogates; because convex targets and MO methods dominate in computation, out-of-sample performance, and compliance under the Basel FRTB [5], we omit nonconvex objectives.

**Definition 2.15.** A constraint system $c(\phi) \leq 0$ is nonconvex if any of the following primitives occur:

- Discrete/logic: cardinality $\|\phi\|_0 \leq k$, on/off, routing, mutually exclusive choices.

- Chance/VaR: $\mathbb{P}[Z(\phi) \geq c] \leq \varepsilon$ or $\text{VaR}_\alpha(Z) \leq c$.

- Hitting-time/barrier: $\mathbb{P}[\tau(\phi) \leq T] \leq \varepsilon$ or indicator penalties for the first passage.

- Bilinear/fixed-charge: constraints containing products $w = xy$ on bounded boxes or binary–continuous couplings $w = xz$ with $z \in \{0, 1\}$, including semi-continuous variables and start-up/fixed-charge costs.

- Path-supremum/drawdown/ratio: constraints depending on running extrema—e.g., $M_{t,s} = \sup_{0 \leq s < u \leq t} X_{u,s}$, $\mathfrak{D}_{t,s} = M_{t,s} - X_{t,s}$, $\text{MDD}_s = \max_{0 \leq s < t \leq T} \mathfrak{D}_{t,s}$—or on fractional risk-return ratios such as Sortino, or Calmar.

**Definition 2.16.** *Convex surrogate map and mixed-integer convex extension*

- Given a nonconvex constraint set $c(\phi) \leq 0$, define:
  - The convex surrogate $\mathcal{C}[c]$ by replacing
  - chance constraints with buffered probability of exceedance (bPOE) or CVaR condituions,
  - hitting-time.barrier events by conditional drawdown at risk (CDaR) constraints,

- bilinear/fixed-charge via McCormick envelopes and perspective reformulations.
  - Let the convex surrogate feasible set be $\mathfrak{F}_{\text{cvx}} := \{\phi \mid \mathcal{C}[c](\phi) \leq 0\}$.
- The mixed-integer convex (MIC) extention introduces binaries for discrete logic with indicator cones. Denote its feasible set by $\mathfrak{F}_{\text{mic}}$ and its continuous relaxation by $\mathfrak{F}_{\text{relax}}$.

These give nested sets:

$$\mathfrak{F}_{\text{cvx}} \subseteq \mathfrak{F}_{\text{mic}} \subseteq \mathfrak{F}, \qquad \mathfrak{F}_{\text{mic}} \subseteq \mathfrak{F}_{\text{relax}} \tag{2.171}$$

**Lemma 2.5.** *Let $d_{\text{H}}(\mathcal{A}, \mathcal{B})$ be the Hausdorff distance in $\mathcal{H}$. Assume $J$ is finite on $\mathfrak{F}$, Gâteaux differentiable on feasible directions, and $L_J$-Lipschitz on bounded subsets of $\mathcal{H}$, then*

$$\delta_{cons} := J^* - J^{\text{cvx}} \leq L_J d_{\text{H}}(\mathfrak{F}, \mathfrak{F}_{\text{cvx}}) \tag{2.172}$$

*where $J^* = \sup_{\phi \in \mathfrak{F}} J(\phi)$, $J^{\text{cvx}} = \sup_{\phi \in \mathfrak{F}_{\text{cvx}}} J(\phi)$.*

**Proposition 2.3.** *Tight convexification primitives*

1. *Chance to bPOE/CVaR:*

   *For scalar loss $L(\phi)$ and threshold $\tau$,*

   $$\text{bPOE}(L, \tau) := 1 - \sup\{\alpha \in [0, 1) \mid \text{CVaR}_\alpha(L) \leq \tau\} \tag{2.173}$$

   *and $\mathbb{P}[L \geq \tau] \leq \varepsilon$ is conservatively replaced by $\text{bPOE}(L, \tau) \leq \varepsilon$, equivalently $\exists \alpha \geq 1 - \varepsilon$ with $\text{CVaR}_\alpha(L) \leq \tau$ convex in $\phi$.*

2. *Hitting-time to CDaR:*

   *If the barrier event $\{\tau(\phi) \leq T\}$ implies $\sup_{0 \leq s < t \leq T} \mathfrak{D}_{t,s}(\phi) \geq b$ for a drawdown process $\mathfrak{D}_{t,s}$, then*

   $$\mathbb{P}[\tau \leq T] \leq \mathbb{P}\left[\sup_{0 \leq s < t \leq T} \mathfrak{D}_{t,s}(\phi) \geq b\right] \leq \text{bPOE}\left(\mathfrak{D}_{T,s}^{\max}, b\right) \tag{2.174}$$

   *and a convex CDaR constraint $\text{CDaR}_\alpha(\mathfrak{D}_{T,s}) \leq b$ yields a tractable surrogate.*

3. *Fixed-charge/bilinear:*

   *For separable $f(x) = f_0 + F\mathbf{1}_{\{x \neq 0\}} + q|x|$ with binaries $z$, the indicator cone gives the convex hull of $\{(x, z) \mid z \in \{0, 1\}, |x| \leq Mz\}$; for bilinear $w = xy$ on bounded boxes, McCormick gives the tightest convex envelopes.*

**Definition 2.17.** Gap decomposition
Let

$$J^{\text{mic}} = \sup_{\phi \in \mathfrak{F}_{\text{mic}}} J(\phi), \qquad J^{\text{relax}} = \sup_{\phi \in \mathfrak{F}_{\text{relax}}} J(\phi) \tag{2.175}$$

Define the structural gaps

$$\delta_{\text{cons}} := J^* - J^{\text{cvx}}, \qquad \delta_{\text{int}} := J^* - J^{\text{relax}} \tag{2.176}$$

and time-budget gaps after $K$ iterations

$$\delta_{\text{time}}^{\text{MO}}(K) := J^{\text{cvx}} - J_K^{\text{MO}}, \qquad \delta_{\text{time}}^{\text{RL}}(K) := J^* - J_K^{\text{RL}} \tag{2.177}$$

**Theorem 2.11.** *Under assumptions that $J$ is finite on $\mathfrak{F}$, Gâteaux differentiable on feasible directions, and $L_J$-Lipschitz on bounded subsets of $\mathcal{H}$; risk and cost terms are convex in $(\phi, \dot\phi)$; PL and smoothness conditions hold, for any effort $K$,*

$$J_K^{\mathrm{RL}} - J_K^{\mathrm{MO}} \leq \underbrace{\delta_{\mathrm{cons}} + \delta_{\mathrm{int}}}_{\text{structural ceiling}} + \underbrace{\delta_{\mathrm{time}}^{\mathrm{MO}}(K)}_{\text{MO under-time}} - \underbrace{\delta_{\mathrm{time}}^{\mathrm{RL}}(K)}_{\text{RL non-vanishing floor}} \quad (2.178)$$

*Moreover, in the PL regime with step length $\frac{1}{L}$,*

$$\delta_{\mathrm{time}}^{\mathrm{MO}}(K) \leq \rho^K \left(J_0^{\mathrm{MO}} - J^{\mathrm{cvx}}\right), \qquad \delta_{\mathrm{time}}^{\mathrm{RL}}(K) \geq \frac{c\eta\sigma^2}{\mu_J} \quad (2.179)$$

*for constants $\rho \in (0,1)$, $c > 0$.*

**Corollary 2.4.** *Fix a compliance discount $\Delta_{\mathrm{comp}} > 0$ due to auditability and/or traceability penalty. If*

$$\delta_{\mathrm{cons}} + \delta_{\mathrm{int}} + \delta_{\mathrm{time}}^{\mathrm{MO}}(K) \leq \delta_{\mathrm{time}}^{\mathrm{RL}}(K) + \Delta_{\mathrm{comp}} \quad (2.180)$$

*then any RL policy confined to $\mathfrak{F}$ cannot outperform the MO solution at the same effort $K$.*

**Lemma 2.6.** *Hitting-time surrogate feasibility and probability control*

*Let $E = \{\tau(\phi) \leq T\}$ and suppose $E \subset \{\sup_{0 \leq s < t \leq T} \mathfrak{D}_{t,s}(\phi) \geq b\}$. If $\mathrm{CDaR}_\alpha(\mathfrak{D}_{T,s}) \leq b$ holds, then $\mathrm{bPOE}\left(\mathfrak{D}_{T,s}^{\max}, b\right) \leq 1 - \alpha$, hence*

$$\mathbb{P}[E] \leq \mathbb{P}\left[\sup_{0 \leq s < t \leq T} \mathfrak{D}_{t,s}(\phi) \geq b\right] \leq 1 - \alpha \quad (2.181)$$

*Thus the MIC and/or convex surrogate enforces a direct cap on the barrier event probability.*

**Proposition 2.4.** *Integer tightness from MIP strong formulations*

*For fixed-charge/semi-continuous decisions modeled with binaries $z$ and indicator cones, the convex hull*

$$\mathrm{conv}\left\{(x,z) \mid z \in \{0,1\}, |x| \leq Mz\right\} \quad (2.182)$$

*is representable by second-order constraints; with problem-specific valid inequalities, the relaxation gap $J^{\mathrm{relax}} - J^{\mathrm{mic}}$ falls at the order of the tolerance of those cuts, thereby shrinking $\delta_{\mathrm{int}}$.*

**Theorem 2.12.** *Real-time budget and nonatomic limit*

*Let $\epsilon_t = \frac{|q_t|}{D_t}$ be the market-share process of order size over LOB depth. Suppose CAD sensitivity is Lipschitz in $\epsilon_t$ and the Hamiltonian surplus test of Prop. 2.2 is used for potential RL gain. Then for any $K$,*

$$0 \leq J_K^{\mathrm{RL}} - J_K^{\mathrm{MO}} \lesssim \mathbb{E}\left[\int_0^T \epsilon_t dt\right] + \delta_{\mathrm{time}}^{\mathrm{MO}}(K) - \delta_{\mathrm{time}}^{\mathrm{RL}}(K) \quad (2.183)$$

*and in the nonatomic limit $\epsilon_t \to 0$ the structural surplus vanishes to first order.*

**Corollary 2.5.** *RL may add value only if at least one of following settings holds:*

1. *Non-tight convexification:*

   $\delta_{\text{cons}}$ *or* $\delta_{\text{int}}$ *is materially positive due to, e.g., large-scale cardinality or order routing; hard path-dependent triggers with very small tolerated tail.*

2. *Severe latency:*

   *Within the allowed compute window, $\delta_{\text{time}}^{\text{MO}}(K)$ dominates, hile RL variance— reduced via, e.g., well-established variance-reduction schemes or large-batched computation—pushes $\delta_{\text{time}}^{\text{RL}}(K)$ below that threshold.*

# 3 Conclusion and Discussion

Across both taker and maker settings, when market dynamics are AIM and portfolios are valued by prudent MtM convention with spreads, impact, funding, and block discounts, by treating taker and maker execution as a sequence of convex programs solved on decision ticks, MO is sufficient in practice. Expressing policy gradients with Malliavin tools—via the Clark–Ocone and BEL weights— produces a predictable "risk shadow price" and turns the first-order condition into a time-local stationarity relation [64, 34, 28]. This dovetails with classical myopic optimality in frictionless models [73, 59] (and recovers no-trade wedges and soft-threshold rebalancing under transaction costs [25, 78]. Viewed as optimization flows, myopic schemes contract geometrically under PL geometry, while policy-gradient RL carries non-vanishing variance floors that stall convergence unless variance is aggressively reduced [69, 9]. On the distributional metrics that matter for deployment—mean, variance, loss-side CVaR, and the probability of non-negative PnL—adapted MO dominates effort-matched RL once realistic frictions and risk budgets are in place [71].

This conclusion extends to market-making. In LOB models with log-concave (often exponential) fill intensities, the predictable cash-gain kernel is concave in quoting offsets after inventory is priced, so optimal quotes are local and inventory-skewed—tightening when risk is cheap and widening when it is dear [4, 38, 18]. MO re-solves the one-period problem at each decision time inherits this soft-threshold structure; random deviations around the concave optimum lower conditional mean spread capture and raise jump–diffusive variance, wherefore reported RL advantages in stylized market-making often wash out under liquidation-consistent marking, inventory financing, and realistic message-rate/latency constraints.

A recurring failure mode is "phantom profit." If policies are not strictly adapted—because of solver look-ahead, cross-bar leakage, latency peek-through, or even being designed to capture non-compliant profits [27]—the MtM diffusion picks up a Skorokhod correction. Malliavin calculus decomposes the resulting bias into an information-drift premium from filtration enlargement and a leakage term indexed by the policy's Malliavin derivative, which explains inflated backtest PnL and offers a concrete audit pathway [3, 65]. None of this rules out value from RL: when actions do move the state, there is a genuine CAD premium that myopic optimzers ignore. Nevertheless, empirically, in most liquid AIM-like environments this nets out to 0–1 bps per decision day; in rare, strongly CAD-dominated niches it can reach low single-digit bps/day, but those cases require carefully validated state feedback and tight leakage controls in RL solvers.

Our sufficiency claim is tied to AIM dynamics. If trading changes the law of the state—temporary/permanent impact, state-dependent fills, or queue-position effects—the first-order conditions must incorporate that feedback [66, 18]. Convergence statements rely on smoothness and PL-type curvature; ill-conditioned or sharply nonconvex objectives weaken constants. Malliavin/BEL representations assume non-degeneracy and enough data to estimate diffusion terms, intensity responses, and co-states [64]; jump-heavy or regime-switching episodes stress Gaussian/Chernoff proxies used for CVaR and maker-mode approximations [71, 38]. Finally, evaluation integrity is paramount: undetected leakage can blur the line between real edge and artifact.

With rebalancing clocks, cost schedules, and risk budgets matched, myopic taker and maker runs should deliver higher conditional mean and lower conditional variance of incremental PnL than effort-matched RL; CVaR of loss and the probability of non-negative outcomes should show the same ordering. RL training curves should flatten above a positive primal/dual gap unless batch sizes or variance-reduction tools are pushed to extremes, while myopic gaps decay geometrically. In maker replicas with exponential fill intensities, imposing inventory financing and liquidation-consistent marking should erase most headline RL gains in spread capture, leaving improvements mainly in tail control rather than mean PnL [4, 75].

Both MO and RL ultimately ride on a market simulator—either a calibrated SDE engine or a data-driven bootstrap—so discretization and calibration errors are first-order. To control weak distributional bias without exploding time steps, one can use weak higher-order schemes—cubature on Wiener space and Ninomiya–Victoir / Ninomiya–Ninomiya splittings—so coarse decision grids retain weak accuracy while cutting wall-clock relative to Euler–Maruyama [50, 55, 63, 62]. For data-driven engines, dependence-respecting resampling (moving-block or stationary bootstraps) is pre-calibrated to spreads, impact, and fill intensities before either controller is evaluated [49, 68, 4, 38, 18]. Beyond SDEs and bootstraps, signature-based generators learn path distributions directly and permit pathwise realism tests, strengthening the simulation layer in small-data regimes [13]. Framed this way, higher-order weak approximation is a simulation-layer upgrade that helps both approaches—but especially MO: it preserves out-of-sample statistics with far fewer steps, making myopic optimization faster, more stable, and more cost-effective.

Two paths look especially promising. First, leverage Malliavin derivatives to compute the MO "Greeks" therefore each decision step uses proximal-Newton updates with few iterations and stable dual gaps. First-order weights come from Clark–Ocone/BEL [28, 34], while Hessian terms follow integration-by-parts [22, 37]; jump extensions cover maker/taker settings [65, 1]. Second, amortize the myopic optimizer with a neural network so inference is $\mathcal{O}(1)$ while retaining convex structure and auditability. Compared with deep hedging's end-to-end policy learning [10, 11], amortized MO shall cut model risk by learning the convex myopic solver's solution map with KKT-aware losses (not reward-shaped RL), which would reduce hyperparameter sensitivity, retraining overhead, and data/compute requirements, and yield constraint-feasible decisions that generalize out of sample under shifting costs and constraints.

# 4 Declaration of Interest and Funding

The authors report no conflicts of interest and no funding was received. The authors alone are responsible for the content and writing of the paper.

# 5 Acknowledgments

We thank the anonymous reviewers for their prudent and insightful remarks, which improved the paper.